\documentclass{aa}
\usepackage{graphicx}
\begin{document}
 
\title{On the metallicity of open clusters I. Photometry}
 
\author{E.~Paunzen\inst{1} 
\and U.~Heiter\inst{2}
\and M.~Netopil\inst{1,3}
\and C.~Soubiran\inst{4}}


\institute{Institut f{\"u}r Astronomie der Universit{\"a}t Wien, T{\"u}rkenschanzstr. 17, A-1180 Wien, Austria
\email{Ernst.Paunzen@univie.ac.at}
\and Department of Physics and Astronomy, Uppsala University, Box 515, SE-75120 Uppsala, Sweden
\and Hvar Observatory, Faculty of Geodesy, University of Zagreb, Ka\v{c}i\'{c}eva 26, HR-10000 Zagreb, Croatia
\and Universit{\'e} Bordeaux 1, CNRS, Laboratoire d'Astrophysique de Bordeaux, BP 89, 33270 Floirac, France}

\date{}

\abstract
{Metallicity is one of four free parameters typically considered when fitting isochrones to the 
cluster sequence. Unfortunately, this parameter is often ignored or assumed to be solar
in most papers. Hence an unknown bias is introduced 
in the estimation of the other three cluster parameters (age, reddening and distance).
Furthermore, studying the metallicity of open clusters allows us not only to derive the Galactic
abundance gradient on a global scale, but also to trace the local solar environment
in more detail.}
{In a series of three papers, we investigate the current status of published 
metallicities for open clusters from widely different photometric and spectroscopic 
methods. A detailed comparison of the results allows us to establish more
reliable photometric calibrations and corrections for isochrone fitting techniques.
Well established databases such as WEBDA help us to perform a homogeneous
analysis of available measurements for a significant number of open clusters.}
{The literature was searched for [Fe/H] estimates on the basis of photometric
calibrations in any available filter system. On the basis of results published by
Tadross,
we demonstrate the caveats of the calibration choice and its possible impact.
In total, we find 406 individual metallicity values for 188 open clusters
within 64 publications. The
values were, finally, unweightedly averaged. }
{Our final sample includes [Fe/H] values for 188 open clusters. Tracing the solar
environment within 4000$\times$4000\,pc$^{2}$ we identify a patchy metallicity distribution as an
extension to the Local Bubble that significantly influences the estimation of the
Galactic metallicity gradient, even on a global scale. In addition, further 
investigations of more distant open clusters are clearly needed to obtain
a more profound picture at Galactocentric distances beyond 10\,000\,pc.}
{Only a combination of all available photometric and spectroscopic data will
shed more light on how the local and global Galactic properties 
are correlated with metallicity. }
\maketitle

\section{Introduction}

One of the most important 
key parameters for our understanding of stellar formation and evolution,
is the intrinsic metallicity of (proto-)stars of a given mass. Even in the
early stages of stellar evolution, the metallicity severely influences 
the cooling and collapse of ionized gas (Jappsen et al. 2007). By comparing detailed
simulations with observations, it has been shown that clouds of lower 
metallicity have a higher probability of fragmentation, indicating that the 
binary frequency is a decreasing function of the cloud metallicity
(Machida 2008). 

Looking at the global properties of our Milky Way, a radial
metallicity gradient throughout the Galactic disk was discovered 
several decades ago, which provides strong
constraints on the mechanism of galaxy formation. Models now show 
that the stellar formation as a function of 
Galactocentric distance strongly influence the appearance and
the development of the metallicity gradients (Chiappini et al. 2001).
Our knowledge is based on stellar data, for example those of Cepheids 
(Cescutti et al. 2007), or open clusters (Chen et al. 2003) as well as
globular clusters (Yong et al. 2008). However, these approaches are still
not satisfactory. All individual stellar estimates are 
limited to the accurate distance estimation of field
stars and the uncertainties in the spectroscopic abundance analysis for very
distant and therefore faint objects. The 
metallicity compilations of open clusters (for example Chen et al. 2003) are 
normally based on inhomogeneous data sets.

In a series of papers, we concentrate on the metallicity of
open clusters using the results of various techniques and methods.
Our final goals are 1) to derive homogeneous metallicities of all
available and published data, 2) to establish a more robust photometric
calibration on the basis of various filter systems, and 3) to investigate
the influence of metallicity on isochrone fitting techniques. 
There have been several studies (e.g., Twarog et al. 1997, Magrini et al. 2009)
of this nature, but none have taken advantage of all available photometric
as well as spectroscopic data.

Besides the investigation of the global Galactic properties, we
are also able to shed more light on the validity of the automatic 
open cluster parameter estimations performed by Kharchenko et al. (2005).
Their estimation of the age, reddening and distance of 650 open clusters
and the follow-up conclusions (Kharchenko et al. 2009) are all based on
solar metallicity. However, the classical technique of isochrone fitting in
various colour-magnitude diagrams, also incorporates the metallicity as an
a-priori free parameter.    

We present our extensive investigation in a series of three papers divided
into a photometric, a spectroscopic, and a ``calibration'' part.

\section{Target selection and literature assessment} \label{ts}

Metallicities, for photometric observations, are often
listed either as [Fe/H] or Z values. If not stated
otherwise, these parameters can be transformed using 
the helium-to-metal enrichment relation
Y\,=\,0.23\,+\,2.25$\times$Z and a solar value Z\,=\,0.019, as given 
in Sect. 2.1 of Girardi et al. (2000).
Almost all available photometric calibrations derive [Fe/H]
as the default standard parameter for the metallicity.
This is because the iron lines, besides hydrogen and
helium lines, dominate the optical spectrum for a wide variety of main 
sequence stars. Therefore, if integrating over one optical filter, the 
abundance of iron can be used as some kind of standard candle.
Another advantage to using this element is that it is normally 
unaffected by the main sequence evolution.

The metallicity determinations obtained on the basis of isochrone fitting only,
were not taken into consideration because the grid of isochrones are
normally only sparsely available for the Z parameter (Schaller et al. 1992).
Therefore, authors only consider whether the isochrones with a Z value 
higher or lower than that of the Sun, fit the observations
more closely. One typical example is the paper by Piatti et al. (2006) who 
published Z\,=\,0.040, corresponding to [Fe/H]\,=\,+0.37\,dex,
for the open cluster NGC~5288. Unfortunately, 
they do not include a corresponding plot with isochrones for the different 
metallicities. 

We searched the literature for metallicity estimates of open cluster members
on the basis of photometric calibrations using WEBDA\footnote{http://www.univie.ac.at/webda}
as a starting point. The apparent double clusters NGC~2451 A/B were excluded from our 
investigation because there is an unsolved and constant confusion about the 
true nature of these aggregates (Platais et al. 2001). 

In total, we found 406 individual metallicity values for 188 open
clusters in 64 publications.
Table \ref{literature} lists the values and the number
of stars used to derive the metallicities, if available, and the employed
photometric filter systems, which are:
\begin{itemize}
\item Caby: Str{\"o}mgren and the Ca filters measuring the Ca\,II HK lines 
(Anthony-Twarog et al. 1991),
\item DDO: David Dunlap Observatory system
(McClure \& van den Bergh 1968),
\item Johnson: $UBVRI$ colors (Bessell 1995) 
\item Str{\"o}mgren: $uvby\beta$ photometry (Str{\"o}mgren 1966),
\item Vilnius: a seven filter system (Smriglio et al. 1990),
\item Washington: a four filter system introduced by Canterna (1976).
\end{itemize}
We note that open clusters are
investigated in all major photometric systems apart from the Geneva one. 
However, in the Geneva 7-color photometric system, blanketing
sensitive indices are available (Paunzen et al. 2005), which could be used as
a metallicity indicator.

The cluster parameters (age, reddening, and distance) were taken from 
Paunzen \& Netopil (2006). The only exceptions are Loden~807, 
NGC~7209, Ruprecht~20, and Ruprecht~32, for which the values were taken 
from the updated list of
Dias et al. (2002) as well as Berkeley~1 and Ruprecht~46, for which 
the parameters from WEBDA were used.
The distance from the Galactic centre was calculated in the standard
way using 8\,kpc for the Sun (Groenewegen et al. 2008). An error propagation
was applied assuming that the uncertainties in the Galactic coordinates for the
individual open clusters are negligible. 

As a next step, we compared the results of two references using identical
data sets, but different photometric calibrations.

\begin{figure}
\begin{center}
\includegraphics[width=85mm]{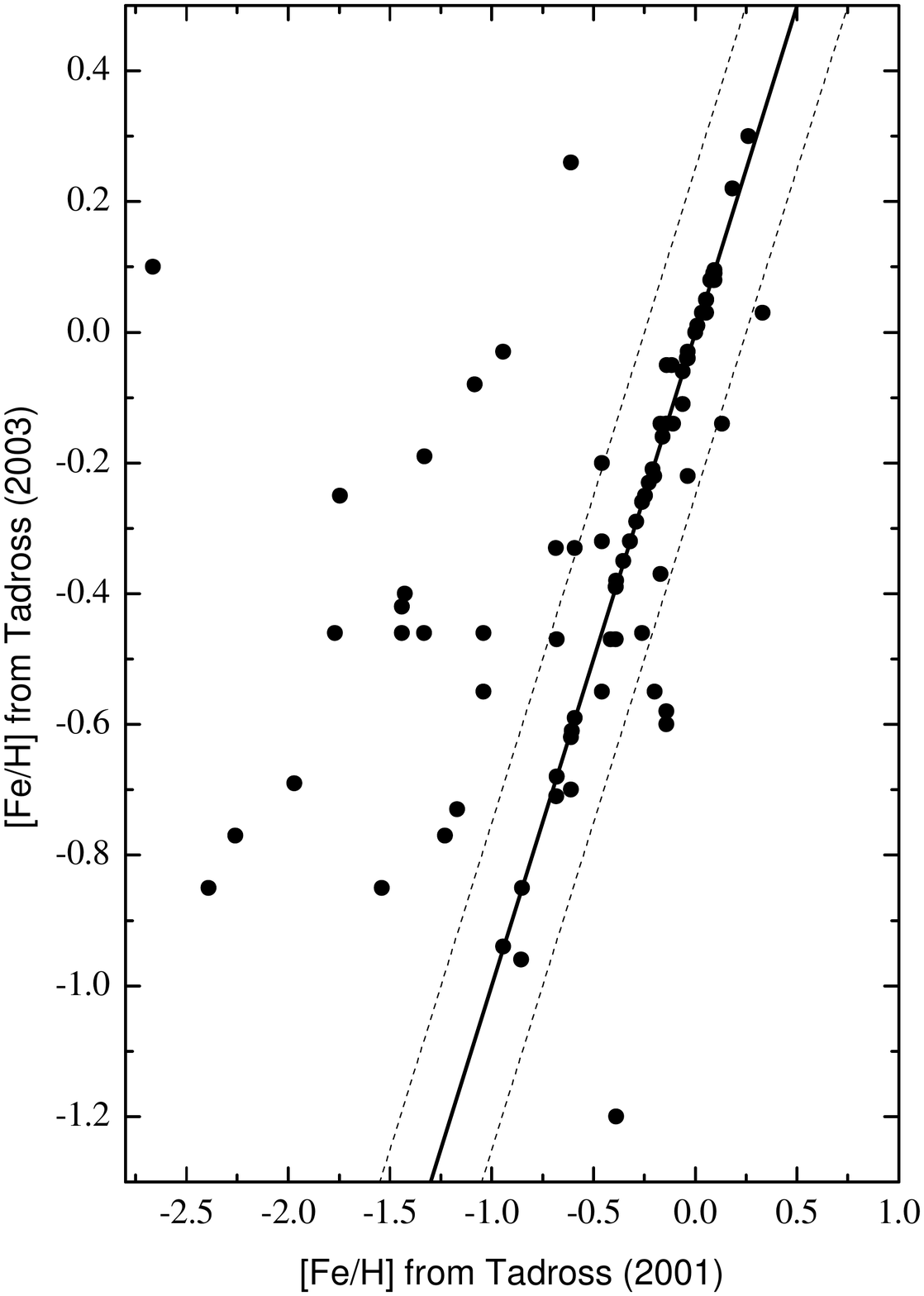}
\caption[]{A comparison of the [Fe/H] values derived via two different photometric
calibrations as published by Tadross (2001, 2003). The dashed lines represent $\pm$0.25\,dex
offsets. Note the wide range of apparent underabundant aggregates in the first paper.}
\label{tadross}
\end{center}
\end{figure}

\begin{table*}
\begin{center}
\caption[]{The [Fe/H] values from photometric calibrations for open clusters found in the literature.}
\begin{tabular}{cccccc|cccccc}
\hline
\hline
Cluster	&	[Fe/H]	&	$\sigma$[Fe/H] &	Stars	&	System	&	Ref	&	Cluster	&	[Fe/H]	&	$\sigma$[Fe/H]	&	Stars	&	
System	&	Ref	\\
\hline
Berkeley 1	&	$-$0.14	&	--	&	--	&	J	&	[57] &	IC 4651	&	+0.18	&	0.05	&	10	&	S	&	[41] \\
Berkeley 2	&	$-$0.59	&	--	&	--	&	J	&	[57] &	IC 4651	&	$-$0.09	&	0.03	&	--	&	J	&	[43] \\
Berkeley 7	&	$-$0.25	&	--	&	--	&	J	&	[57] &	IC 4651	&	+0.08	&	0.09	&	8	&	D	&	[49] \\
Berkeley 14	&	$-$1.20	&	--	&	--	&	J	&	[57] &	IC 4651	&	$-$0.14	&	--	&	--	&	J	&	[57] \\
Berkeley 21	&	$-$0.96	&	--	&	--	&	J	&	[57] &	IC 4651	&	+0.095	&	0.053	&	11	&	D	&	[61] \\
Berkeley 22	&	$-$0.42	&	--	&	7	&	J	&	[43] &	IC 4665	&	$-$0.11	&	0.07	&	5	&	S	&	[40] \\
Berkeley 28	&	$-$0.61	&	--	&	--	&	J	&	[57] &	IC 4725	&	$-$0.092	&	--	&	63	&	J	&	[7] \\
Berkeley 29	&	$-$0.30	&	--	&	19	&	J	&	[43] &	IC 4756	&	+0.003	&	--	&	36	&	J	&	[7] \\
Berkeley 30	&	+0.10	&	--	&	--	&	J	&	[57] &	IC 4756	&	+0.04	&	0.05	&	10	&	D	&	[54] \\
Berkeley 31	&	$-$0.60	&	--	&	--	&	J	&	[57] &	IC 4756	&	$-$0.012	&	0.072	&	6	&	D	&	[61] \\
Berkeley 32	&	$-$0.37	&	0.05	&	--	&	J, W	&	[32] &	IC 4996	&	$-$0.47	&	--	&	--	&	J	&	[57] \\
Berkeley 32	&	$-$0.37	&	0.04	&	--	&	J	&	[43] &	King 2	&	$-$0.32	&	--	&	--	&	J	&	[57] \\
Berkeley 32	&	$-$0.55	&	--	&	--	&	J	&	[57] &	King 7	&	+0.03	&	--	&	--	&	J	&	[57] \\
Berkeley 39	&	$-$0.42	&	0.11	&	8	&	W	&	[29] &	King 10	&	+0.09	&	--	&	--	&	J	&	[57] \\
Berkeley 39	&	$-$0.26	&	--	&	27	&	J	&	[43] &	King 11	&	$-$0.31	&	--	&	21	&	J	&	[43] \\
Berkeley 39	&	$-$0.33	&	--	&	--	&	J	&	[57] &	King 11	&	$-$0.37	&	--	&	--	&	J	&	[57] \\
Berkeley 42	&	+0.08	&	--	&	--	&	J	&	[57] &	Loden 807	&	$-$0.19	&	0.06	&	2	&	D	&	[25] \\
Berkeley 58	&	+0.09	&	--	&	--	&	J	&	[57] &	Loden 807	&	$-$0.22	&	--	&	1	&	D	&	[49] \\
Berkeley 60	&	$-$0.77	&	--	&	--	&	J	&	[57] &	Loden 807	&	$-$0.130	&	0.090	&	1	&	D	&	[61] \\
Berkeley 62	&	$-$0.85	&	--	&	--	&	J	&	[57] &	Melotte 20	&	+0.07	&	--	&	119	&	J	&	[7] \\
Berkeley 79	&	$-$0.32	&	--	&	--	&	J	&	[57] &	Melotte 20	&	+0.04	&	0.02	&	19	&	S	&	[40] \\
Berkeley 86	&	$-$0.08	&	--	&	--	&	J	&	[57] &	Melotte 20	&	+0.01	&	0.09	&	19	&	S	&	[41] \\
Collinder 74	&	+0.05	&	--	&	--	&	J	&	[57] &	Melotte 22	&	+0.08	&	--	&	31	&	J	&	[7] \\
Collinder 140	&	$-$0.092	&	--	&	21	&	J	&	[7] &	Melotte 22	&	+0.05	&	0.02	&	30	&	S	&	[40] \\
Collinder 140	&	+0.00	&	0.20	&	--	&	S	&	[37] &	Melotte 22	&	+0.06	&	0.12	&	23	&	S	&	[41] \\
Collinder 173	&	$-$0.06	&	0.20	&	--	&	S	&	[37] &	Melotte 25	&	+0.08	&	--	&	48	&	J	&	[7] \\
Collinder 258	&	+0.07	&	0.21	&	2	&	D	&	[17] &	Melotte 25	&	+0.13	&	0.01	&	49	&	S	&	[40] \\
Collinder 258	&	$-$0.24	&	--	&	1	&	D	&	[49] &	Melotte 25	&	+0.12	&	0.09	&	42	&	S	&	[41] \\
Collinder 258	&	$-$0.090	&	0.090	&	1	&	D	&	[61] &	Melotte 25	&	+0.19	&	0.03	&	4	&	D	&	[49] \\
Collinder 272	&	$-$0.40	&	--	&	--	&	J	&	[57] &	Melotte 25	&	+0.175	&	0.034	&	4	&	D	&	[61] \\
Haffner 8	&	$-$0.09	&	0.10	&	2	&	D, W	&	[17] &	Melotte 66	&	$-$0.35	&	0.20	&	4	&	W	&	[28] \\
Haffner 8	&	+0.06	&	0.06	&	2	&	D	&	[49] &	Melotte 66	&	$-$0.52	&	0.05	&	--	&	J	&	[43] \\
Haffner 8	&	+0.060	&	0.040	&	2	&	D	&	[61] &	Melotte 66	&	$-$0.44	&	0.09	&	6	&	D	&	[49] \\
IC 348	&	+0.03	&	--	&	--	&	J	&	[57] &	Melotte 66	&	$-$0.53	&	0.08	&	--	&	S	&	[60] \\
IC 1311	&	$-$0.23	&	--	&	--	&	J	&	[57] &	Melotte 66	&	$-$0.349	&	0.145	&	7	&	D	&	[61] \\
IC 1590	&	$-$0.73	&	--	&	--	&	J	&	[57] &	Melotte 71	&	$-$0.57	&	0.18	&	14	&	W	&	[29] \\
IC 1805	&	+0.08	&	--	&	--	&	J	&	[57] &	Melotte 71	&	$-$0.17	&	0.02	&	--	&	C, S	&	[62] \\
IC 2391	&	$-$0.15	&	--	&	--	&	S	&	[27] &	Melotte 105	&	+0.00	&	--	&	--	&	J	&	[57] \\
IC 2391	&	$-$0.09	&	0.20	&	--	&	S	&	[37] &	Melotte 111	&	$-$0.063	&	--	&	30	&	J	&	[7] \\
IC 2488	&	+0.095	&	0.08	&	3	&	D, W	&	[22] &	Melotte 111	&	$-$0.03	&	0.02	&	18	&	S	&	[40] \\
IC 2602	&	$-$0.232	&	--	&	--	&	J	&	[7] &	Melotte 111	&	$-$0.01	&	0.07	&	17	&	S	&	[41] \\
IC 2714	&	$-$0.12	&	0.09	&	19	&	D, W	&	[19] &	NGC 103	&	$-$0.85	&	--	&	--	&	J	&	[57] \\
IC 2714	&	+0.01	&	0.10	&	10	&	D	&	[49] &	NGC 129	&	$-$0.46	&	--	&	--	&	J	&	[57] \\
IC 2714	&	$-$0.011	&	0.116	&	5	&	D	&	[61] &	NGC 188	&	+0.055	&	--	&	34	&	J	&	[7] \\
IC 4651	&	+0.077	&	0.012	&	9	&	S	&	[1] &	NGC 188	&	$-$0.16	&	0.11	&	7	&	D	&	[49] \\
IC 4651	&	$-$0.194	&	--	&	96	&	J	&	[7] &	NGC 188	&	$-$0.046	&	0.098	&	8	&	D	&	[61] \\
IC 4651	&	+0.21	&	0.03	&	31	&	D	&	[10] &	NGC 366	&	$-$0.55	&	--	&	--	&	J	&	[57] \\
IC 4651	&	+0.15	&	0.15	&	--	&	W	&	[10] &	NGC 381	&	$-$0.04	&	--	&	--	&	J	&	[57] \\
IC 4651	&	+0.11	&	0.05	&	--	&	D, W	&	[17] &	NGC 433	&	$-$0.68	&	--	&	--	&	J	&	[57] \\
\hline
\multicolumn{12}{l}{{\tiny Photometric systems: Caby (C) DDO (D) Johnson (J) 
Str{\"o}mgren (S) Vilnius (V) Washington (W)}} \\
\multicolumn{12}{l}{{\tiny Literature: [1] Anthony-Twarog \& Twarog (2000) [2] 
Anthony-Twarog \& Twarog (2004) [3] Anthony-Twarog \& Twarog (2006)}} \\
\multicolumn{12}{l}{{\tiny [4] Anthony-Twarog et al. (2005) [5] Anthony-Twarog et al. (2006) 
[6] Bruntt et al. (1999) [7] Cameron (1985b) [8] Clari{\'a} (1982)}} \\
\multicolumn{12}{l}{{\tiny [9] Clari{\'a} (1985) [10] Clari{\'a} \& Lapasset (1983)
[11] Clari{\'a} \& Lapasset (1985) [12] Clari{\'a} \& Lapasset (1986a)}} \\
\multicolumn{12}{l}{{\tiny [13] Clari{\'a} \& Lapasset (1986b) [14] Clari{\'a} \& Lapasset (1989)
[15] Claria \& Mermilliod (1992) [16] Clari{\'a} \& Minniti (1988)}} \\ 
\multicolumn{12}{l}{{\tiny [17] Clari{\'a} et al. (1989) [18] Clari{\'a} et al. (1991) [19] Clari{\'a} et al. (1994)
[20] Clari{\'a} et al. (1996) [21] Clari{\'a} et al. (1999) }} \\ 
\multicolumn{12}{l}{{\tiny [22] Clari{\'a} et al. (2003) [23] Clari{\'a} et al. (2005) [24] Clari{\'a} et al. (2007) 
[25] Clari{\'a} et al. (2008) [26] Dawson (1981)}} \\
\multicolumn{12}{l}{{\tiny [27] Eggen (1983) [28] Geisler \& Smith (1984)
[29] Geisler et al. (1992) [30] Janes (1984) [31] Janes \& Smith (1984) }} \\
\multicolumn{12}{l}{{\tiny [32] Kaluzny \& Mazur (1991a) [33] Kaluzny \& Mazur (1991b)
[34] Kaluzny \& Mazur (1991c) [35] Kaluzny \& Rucinski (1995)}} \\
\multicolumn{12}{l}{{\tiny [36] Kyeong et al. (2008) [37] Lynga \& Wramdemark (1984)
[38] McClure et al. (1981) [39] Mermilliod et al. (2001)}} \\
\multicolumn{12}{l}{{\tiny [40] Nissen (1980) [41] Nissen (1988)
[42] Nissen et al. (1987) [43] Noriega-Mendoza \& Ruelas-Mayorgo (1997)}} \\ 
\multicolumn{12}{l}{{\tiny  [44] Palous \& Hauck (1986)
[45] Parisi et al. (2005) [46] Pastoriza \& Ropke (1983) [47] Paunzen et al. (2003)}} \\
\multicolumn{12}{l}{{\tiny [48] Philip (1976) [49] Piatti et al. (1995)
[50] Piatti et al. (2003a) [51] Piatti et al. (2003b) [52] Piatti et al. (2004)}} \\
\multicolumn{12}{l}{{\tiny  [53] Richtler (1985) [54] Smith (1983)
[55] Smith \& Hesser (1983) [56] Sung et al. (2002) [57] Tadross (2003) }} \\
\multicolumn{12}{l}{{\tiny  [58] Twarog (1983)
[59] Twarog et al. (1993) [60] Twarog et al. (1995) [61] Twarog et al. (1997) [62] Twarog et al. (2006)}} \\
\multicolumn{12}{l}{{\tiny [63] Twarog et al. (2007)
[64] Vansevicius et al. (1997)}} \\
\end{tabular}
\label{literature}
\end{center}
\end{table*}

\begin{table*}
\begin{center}
\addtocounter{table}{-1}
\caption[]{continued}
\begin{tabular}{cccccc|cccccc}
\hline
\hline
Cluster	&	[Fe/H]	&	$\sigma$[Fe/H] &	Stars	&	System	&	Ref	&	Cluster	&	[Fe/H]	&	$\sigma$[Fe/H]	&	Stars	&	
System	&	Ref	\\
\hline
NGC 436	&	$-$0.77	&	--	&	--	&	J	&	[57] &	NGC 2301	&	+0.060	&	0.060	&	2	&	D	&	[61] \\
NGC 457	&	$-$0.46	&	--	&	--	&	J	&	[57] &	NGC 2324	&	$-$0.163	&	--	&	64	&	J	&	[7] \\
NGC 581	&	$-$0.85	&	--	&	--	&	J	&	[57] &	NGC 2324	&	$-$1.01	&	0.27	&	9	&	W	&	[29] \\
NGC 654	&	$-$0.68	&	--	&	--	&	J	&	[57] &	NGC 2324	&	$-$0.31	&	0.04	&	5	&	W	&	[52] \\
NGC 663	&	$-$0.70	&	--	&	--	&	J	&	[57] &	NGC 2335	&	+0.22	&	--	&	--	&	D	&	[9] \\
NGC 752	&	$-$0.06	&	0.03	&	10	&	C	&	[3] &	NGC 2335	&	+0.02	&	--	&	1	&	D	&	[49] \\
NGC 752	&	$-$0.311	&	--	&	14	&	J	&	[7] &	NGC 2335	&	$-$0.030	&	0.090	&	1	&	D	&	[61] \\
NGC 752	&	$-$0.26	&	0.06	&	7	&	D	&	[20] &	NGC 2343	&	$-$0.15	&	--	&	--	&	D	&	[9] \\
NGC 752	&	$-$0.02	&	0.02	&	29	&	S	&	[40] &	NGC 2343	&	$-$0.39	&	--	&	1	&	D	&	[49] \\
NGC 752	&	$-$0.05	&	0.13	&	26	&	S	&	[41] &	NGC 2354	&	$-$0.30	&	0.02	&	12	&	D, J, W	&	[21] \\
NGC 752	&	$-$0.22	&	0.05	&	7	&	D	&	[49] &	NGC 2355	&	+0.13	&	--	&	50	&	J	&	[34] \\
NGC 752	&	$-$0.21	&	0.09	&	19	&	D, J, S	&	[58] &	NGC 2355	&	$-$0.14	&	--	&	--	&	J	&	[57] \\
NGC 752	&	$-$0.160	&	0.040	&	6	&	D	&	[61] &	NGC 2360	&	$-$0.092	&	--	&	34	&	J	&	[7] \\
NGC 1039	&	$-$0.291	&	--	&	36	&	J	&	[7] &	NGC 2360	&	$-$0.12	&	0.03	&	13	&	D	&	[25] \\
NGC 1193	&	$-$0.45	&	0.12	&	32	&	J	&	[36] &	NGC 2360	&	$-$0.29	&	0.04	&	6	&	W	&	[29] \\
NGC 1193	&	$-$0.54	&	--	&	16	&	J	&	[43] &	NGC 2360	&	$-$0.14	&	0.07	&	10	&	D	&	[49] \\
NGC 1342	&	$-$0.163	&	--	&	14	&	J	&	[7] &	NGC 2360	&	$-$0.154	&	0.113	&	11	&	D	&	[61] \\
NGC 1342	&	$-$0.24	&	0.12	&	3	&	D	&	[49] &	NGC 2420	&	$-$0.37	&	0.05	&	106	&	C, S	&	[5] \\
NGC 1528	&	$-$0.133	&	--	&	17	&	J	&	[7] &	NGC 2420	&	$-$0.43	&	0.07	&	--	&	J	&	[43] \\
NGC 1545	&	$-$0.16	&	--	&	1	&	D	&	[20] &	NGC 2420	&	$-$0.43	&	0.10	&	8	&	D	&	[49] \\
NGC 1545	&	$-$0.18	&	--	&	1	&	D	&	[49] &	NGC 2420	&	$-$0.281	&	0.073	&	10	&	D	&	[61] \\
NGC 1545	&	$-$0.060	&	0.090	&	1	&	D	&	[61] &	NGC 2422	&	+0.11	&	0.10	&	11	&	S	&	[41] \\
NGC 1662	&	$-$0.232	&	--	&	20	&	J	&	[7] &	NGC 2423	&	+0.00	&	0.08	&	4	&	D, W	&	[17] \\
NGC 1662	&	$-$0.13	&	0.02	&	2	&	D	&	[20] &	NGC 2423	&	+0.03	&	0.03	&	8	&	D	&	[25] \\
NGC 1662	&	$-$0.03	&	0.01	&	2	&	D	&	[49] &	NGC 2423	&	+0.11	&	0.04	&	7	&	D	&	[49] \\
NGC 1662	&	$-$0.095	&	0.015	&	2	&	D	&	[61] &	NGC 2423	&	+0.143	&	0.092	&	3	&	D	&	[61] \\
NGC 1750	&	$-$0.69	&	--	&	--	&	J	&	[57] &	NGC 2437	&	$-$0.16	&	0.11	&	2	&	D	&	[25] \\
NGC 1798	&	$-$0.46	&	--	&	--	&	J	&	[57] &	NGC 2437	&	+0.07	&	--	&	1	&	D	&	[49] \\
NGC 1817	&	$-$0.291	&	--	&	17	&	J	&	[7] &	NGC 2447	&	$-$0.10	&	0.08	&	9	&	D, J, W	&	[23] \\
NGC 1817	&	$-$0.33	&	0.08	&	15	&	W	&	[45] &	NGC 2451	&	$-$0.50	&	0.02	&	--	&	S	&	[37] \\
NGC 1931	&	$-$0.06	&	--	&	--	&	J	&	[57] &	NGC 2451	&	$-$0.106	&	0.122	&	--	&	D	&	[46] \\
NGC 2099	&	+0.18	&	0.07	&	3	&	D	&	[49] &	NGC 2477	&	$-$0.008	&	--	&	29	&	J	&	[7] \\
NGC 2099	&	+0.089	&	0.146	&	5	&	D	&	[61] &	NGC 2477	&	$-$0.13	&	0.18	&	16	&	W	&	[29] \\
NGC 2112	&	$-$1.30	&	0.22	&	4	&	S	&	[53] &	NGC 2477	&	+0.04	&	0.01	&	--	&	D	&	[55] \\
NGC 2158	&	$-$0.36	&	0.18	&	3	&	D	&	[49] &	NGC 2477	&	$-$0.05	&	--	&	--	&	J	&	[57] \\
NGC 2158	&	$-$0.238	&	0.064	&	5	&	D	&	[61] &	NGC 2482	&	+0.20	&	0.01	&	--	&	D	&	[10] \\
NGC 2168	&	$-$0.28	&	--	&	1	&	D	&	[49] &	NGC 2482	&	+0.10	&	0.10	&	--	&	W	&	[10] \\
NGC 2168	&	$-$0.39	&	--	&	--	&	J	&	[57] &	NGC 2482	&	+0.07	&	0.07	&	3	&	D, W	&	[17] \\
NGC 2168	&	$-$0.160	&	0.090	&	1	&	D	&	[61] &	NGC 2482	&	+0.14	&	0.04	&	3	&	D	&	[25] \\
NGC 2192	&	$-$0.22	&	--	&	--	&	J	&	[57] &	NGC 2482	&	+0.13	&	0.04	&	3	&	D	&	[49] \\
NGC 2194	&	$-$0.27	&	0.06	&	9	&	W	&	[50] &	NGC 2482	&	+0.120	&	0.026	&	3	&	D	&	[61] \\
NGC 2204	&	$-$0.618	&	--	&	9	&	J	&	[7] &	NGC 2489	&	+0.00	&	--	&	1	&	D	&	[20] \\
NGC 2204	&	$-$0.41	&	0.19	&	6	&	D	&	[26] &	NGC 2489	&	+0.10	&	--	&	1	&	D	&	[49] \\
NGC 2204	&	$-$0.39	&	0.22	&	5	&	D	&	[49] &	NGC 2489	&	+0.01	&	--	&	--	&	J	&	[57] \\
NGC 2204	&	$-$0.338	&	0.250	&	5	&	D	&	[61] &	NGC 2489	&	+0.080	&	0.010	&	2	&	D	&	[61] \\
NGC 2232	&	$-$0.205	&	0.213	&	--	&	D	&	[46] &	NGC 2506	&	$-$0.58	&	0.14	&	23	&	W	&	[29] \\
NGC 2236	&	$-$0.30	&	0.20	&	13	&	W	&	[24] &	NGC 2506	&	$-$0.57	&	--	&	16	&	D	&	[38] \\
NGC 2243	&	$-$0.57	&	0.03	&	100	&	C, S	&	[4] &	NGC 2506	&	$-$0.48	&	0.08	&	4	&	D	&	[49] \\
NGC 2243	&	$-$0.65	&	0.12	&	4	&	D	&	[49] &	NGC 2506	&	$-$0.58	&	--	&	--	&	J	&	[57] \\
NGC 2243	&	$-$0.480	&	0.160	&	5	&	D	&	[61] &	NGC 2506	&	$-$0.368	&	0.108	&	5	&	D	&	[61] \\
NGC 2244	&	$-$0.46	&	--	&	--	&	J	&	[57] &	NGC 2516	&	$-$0.422	&	--	&	67	&	J	&	[7] \\
NGC 2251	&	$-$0.25	&	0.04	&	3	&	W	&	[45] &	NGC 2516	&	+0.00	&	0.10	&	2	&	D	&	[17] \\
NGC 2251	&	$-$0.17	&	0.06	&	3	&	D	&	[49] &	NGC 2516	&	$-$0.28	&	0.20	&	--	&	S	&	[37] \\
NGC 2264	&	+0.00	&	--	&	--	&	D	&	[9] &	NGC 2516	&	+0.06	&	0.06	&	8	&	S	&	[41] \\
NGC 2264	&	$-$0.09	&	0.30	&	--	&	S	&	[48] &	NGC 2516	&	+0.02	&	--	&	1	&	D	&	[49] \\
NGC 2264	&	$-$0.16	&	--	&	--	&	J	&	[57] &	NGC 2516	&	$-$0.10	&	0.04	&	--	&	J	&	[56] \\
NGC 2266	&	$-$0.26	&	0.20	&	9	&	W	&	[33] &	NGC 2516	&	+0.060	&	0.030	&	2	&	D	&	[61] \\
NGC 2281	&	$-$0.074	&	--	&	38	&	J	&	[7] &	NGC 2527	&	$-$0.02	&	--	&	--	&	D	&	[9] \\
NGC 2287	&	+0.065	&	--	&	14	&	J	&	[7] &	NGC 2527	&	$-$0.09	&	--	&	1	&	D	&	[49] \\
NGC 2287	&	$-$0.10	&	0.11	&	10	&	S	&	[41] &	NGC 2527	&	$-$0.080	&	0.090	&	1	&	D	&	[61] \\
NGC 2287	&	$-$0.246	&	0.255	&	--	&	D	&	[46] &	NGC 2539	&	+0.24	&	0.06	&	--	&	D	&	[12] \\
NGC 2287	&	$-$0.13	&	0.06	&	3	&	D	&	[49] &	NGC 2539	&	+0.00	&	0.20	&	--	&	W	&	[12] \\
NGC 2287	&	+0.040	&	0.022	&	4	&	D	&	[61] &	NGC 2539	&	+0.03	&	0.09	&	7	&	D, W	&	[17] \\
NGC 2301	&	+0.014	&	--	&	18	&	J	&	[7] &	NGC 2539	&	+0.17	&	0.08	&	3	&	D	&	[49] \\
NGC 2301	&	+0.10	&	0.05	&	2	&	D	&	[17] &	NGC 2539	&	+0.137	&	0.062	&	6	&	D	&	[61] \\
NGC 2301	&	+0.04	&	0.12	&	13	&	S	&	[41] &	NGC 2546	&	+0.26	&	--	&	--	&	D	&	[9] \\
NGC 2301	&	+0.01	&	--	&	1	&	D	&	[49] &	NGC 2546	&	+0.11	&	+0.09	&	2	&	D	&	[25] \\
\hline
\end{tabular}
\end{center}
\end{table*}

\begin{table*}
\begin{center}
\addtocounter{table}{-1}
\caption[]{continued}
\begin{tabular}{cccccc|cccccc}
\hline
\hline
Cluster	&	[Fe/H]	&	$\sigma$[Fe/H] &	Stars	&	System	&	Ref	&	Cluster	&	[Fe/H]	&	$\sigma$[Fe/H]	&	Stars	&	
System	&	Ref	\\
\hline
NGC 2546	&	$-$0.14	&	--	&	1	&	D	&	[49] &	NGC 4755	&	$-$0.21	&	--	&	--	&	J	&	[57] \\
NGC 2546	&	+0.120	&	0.130	&	2	&	D	&	[61] &	NGC 4815	&	$-$0.19	&	--	&	--	&	J	&	[57] \\
NGC 2547	&	$-$0.121	&	--	&	43	&	J	&	[7] &	NGC 5138	&	+0.18	&	--	&	1	&	D	&	[49] \\
NGC 2547	&	$-$0.13	&	--	&	1	&	D	&	[8] &	NGC 5138	&	+0.120	&	0.050	&	2	&	D	&	[61] \\
NGC 2547	&	$-$0.18	&	0.20	&	--	&	S	&	[37] &	NGC 5168	&	+0.05	&	--	&	--	&	J	&	[57] \\
NGC 2547	&	$-$0.21	&	--	&	1	&	D	&	[49] &	NGC 5281	&	$-$0.18	&	--	&	1	&	D	&	[17] \\
NGC 2547	&	$-$0.160	&	0.090	&	1	&	D	&	[61] &	NGC 5316	&	+0.19	&	0.11	&	--	&	D, J, W	&	[14] \\
NGC 2548	&	+0.14	&	--	&	--	&	D	&	[9] &	NGC 5316	&	$-$0.02	&	0.12	&	2	&	D	&	[49] \\
NGC 2548	&	+0.01	&	0.02	&	2	&	D	&	[49] &	NGC 5316	&	+0.128	&	0.127	&	4	&	D	&	[61] \\
NGC 2548	&	+0.080	&	0.020	&	3	&	D	&	[61] &	NGC 5606	&	+0.095	&	--	&	--	&	J	&	[57] \\
NGC 2567	&	+0.00	&	--	&	--	&	D	&	[13] &	NGC 5617	&	+0.31	&	0.10	&	2	&	D, W	&	[17] \\
NGC 2567	&	+0.00	&	0.10	&	--	&	W	&	[13] &	NGC 5617	&	$-$0.32	&	--	&	--	&	J	&	[57] \\
NGC 2567	&	$-$0.09	&	0.19	&	2	&	D	&	[49] &	NGC 5662	&	$-$0.03	&	0.13	&	2	&	D, W	&	[18] \\
NGC 2567	&	+0.22	&	--	&	--	&	J	&	[57] &	NGC 5662	&	+0.09	&	--	&	--	&	J	&	[57] \\
NGC 2567	&	$-$0.030	&	0.120	&	2	&	D	&	[61] &	NGC 5822	&	+0.01	&	0.04	&	--	&	D	&	[11] \\
NGC 2571	&	+0.05	&	--	&	23	&	J	&	[7] &	NGC 5822	&	$-$0.25	&	0.15	&	--	&	W	&	[11] \\
NGC 2627	&	$-$0.12	&	0.08	&	8	&	W	&	[51] &	NGC 5822	&	$-$0.12	&	0.08	&	14	&	D, W	&	[17] \\
NGC 2627	&	$-$0.04	&	--	&	--	&	J	&	[57] &	NGC 5822	&	+0.03	&	0.09	&	13	&	D	&	[49] \\
NGC 2632	&	+0.04	&	--	&	76	&	J	&	[7] &	NGC 5822	&	$-$0.15	&	0.015	&	--	&	D, J	&	[59] \\
NGC 2632	&	+0.20	&	0.07	&	3	&	D	&	[20] &	NGC 5822	&	$-$0.020	&	0.086	&	14	&	D	&	[61] \\
NGC 2632	&	+0.09	&	0.02	&	47	&	S	&	[40] &	NGC 6025	&	+0.195	&	--	&	48	&	J	&	[7] \\
NGC 2632	&	+0.10	&	0.15	&	42	&	S	&	[41] &	NGC 6067	&	+0.10	&	0.06	&	6	&	D, W	&	[17] \\
NGC 2632	&	+0.19	&	0.06	&	3	&	D	&	[49] &	NGC 6067	&	$-$0.01	&	0.07	&	3	&	D	&	[49] \\
NGC 2632	&	+0.142	&	0.069	&	4	&	D	&	[61] &	NGC 6067	&	+0.138	&	0.064	&	5	&	D	&	[61] \\
NGC 2658	&	$-$0.94	&	--	&	--	&	J	&	[57] &	NGC 6087	&	$-$0.04	&	--	&	--	&	J	&	[57] \\
NGC 2660	&	$-$1.05	&	0.16	&	7	&	W	&	[29] &	NGC 6134	&	+0.28	&	0.02	&	--	&	S	&	[6] \\
NGC 2660	&	$-$0.27	&	0.13	&	5	&	D	&	[49] &	NGC 6134	&	$-$0.05	&	0.12	&	17	&	D, J, W	&	[15] \\
NGC 2660	&	$-$0.05	&	--	&	--	&	J	&	[57] &	NGC 6134	&	+0.20	&	0.10	&	7	&	D	&	[49] \\
NGC 2660	&	$-$0.181	&	0.118	&	5	&	D	&	[61] &	NGC 6134	&	+0.30	&	--	&	--	&	J	&	[57] \\
NGC 2669	&	$-$0.20	&	--	&	1	&	D	&	[17] &	NGC 6134	&	+0.182	&	0.087	&	10	&	D	&	[61] \\
NGC 2682	&	$-$0.029	&	--	&	58	&	J	&	[7] &	NGC 6192	&	$-$0.10	&	0.09	&	--	&	S	&	[47] \\
NGC 2682	&	$-$0.05	&	0.03	&	22	&	D	&	[31] &	NGC 6192	&	$-$0.35	&	--	&	--	&	J	&	[57] \\
NGC 2682	&	$-$0.06	&	0.07	&	36	&	S	&	[42] &	NGC 6204	&	$-$0.14	&	--	&	--	&	J	&	[57] \\
NGC 2682	&	$-$0.05	&	0.04	&	--	&	J	&	[43] &	NGC 6208	&	$-$0.03	&	0.06	&	3	&	D	&	[49] \\
NGC 2682	&	$-$0.01	&	0.11	&	14	&	D	&	[49] &	NGC 6231	&	+0.26	&	--	&	--	&	J	&	[57] \\
NGC 2682	&	$-$0.11	&	--	&	--	&	J	&	[57] &	NGC 6259	&	+0.06	&	0.08	&	8	&	W	&	[39] \\
NGC 2682	&	+0.000	&	0.092	&	25	&	D	&	[61] &	NGC 6259	&	+0.01	&	0.09	&	2	&	D	&	[49] \\
NGC 2818	&	$-$0.31	&	0.13	&	11	&	W	&	[29] &	NGC 6281	&	+0.07	&	--	&	37	&	J	&	[7] \\
NGC 2818	&	$-$0.29	&	--	&	--	&	J	&	[57] &	NGC 6281	&	$-$0.08	&	0.08	&	2	&	D, W	&	[17] \\
NGC 2910	&	$-$0.04	&	--	&	--	&	J	&	[57] &	NGC 6281	&	+0.00	&	0.12	&	2	&	D	&	[49] \\
NGC 2972	&	$-$0.15	&	0.08	&	2	&	D	&	[17] &	NGC 6281	&	+0.005	&	0.095	&	2	&	D	&	[61] \\
NGC 2972	&	$-$0.09	&	0.01	&	2	&	D	&	[49] &	NGC 6405	&	+0.07	&	--	&	20	&	J	&	[7] \\
NGC 2972	&	$-$0.073	&	0.021	&	3	&	D	&	[61] &	NGC 6425	&	+0.25	&	0.04	&	2	&	D	&	[25] \\
NGC 3114	&	$-$0.04	&	0.04	&	7	&	D, W	&	[17] &	NGC 6425	&	+0.09	&	--	&	1	&	D	&	[49] \\
NGC 3114	&	$-$0.14	&	0.14	&	4	&	D	&	[49] &	NGC 6451	&	$-$0.34	&	0.08	&	--	&	S	&	[47] \\
NGC 3114	&	+0.022	&	0.070	&	5	&	D	&	[61] &	NGC 6451	&	$-$0.20	&	--	&	--	&	J	&	[57] \\
NGC 3532	&	+0.08	&	0.08	&	5	&	W	&	[16] &	NGC 6475	&	+0.02	&	0.07	&	11	&	S	&	[41] \\
NGC 3532	&	$-$0.10	&	0.09	&	3	&	D	&	[49] &	NGC 6475	&	+0.03	&	--	&	1	&	D	&	[49] \\
NGC 3532	&	$-$0.022	&	0.088	&	5	&	D	&	[61] &	NGC 6475	&	+0.070	&	0.090	&	1	&	D	&	[61] \\
NGC 3680	&	$-$0.14	&	0.03	&	30	&	C, S	&	[2] &	NGC 6494	&	$-$0.14	&	0.15	&	--	&	D, J, W	&	[14] \\
NGC 3680	&	+0.09	&	0.02	&	--	&	S	&	[6] &	NGC 6494	&	+0.13	&	0.15	&	2	&	D	&	[49] \\
NGC 3680	&	$-$0.145	&	--	&	27	&	J	&	[7] &	NGC 6494	&	+0.090	&	0.117	&	4	&	D	&	[61] \\
NGC 3680	&	+0.00	&	0.04	&	--	&	D	&	[10] &	NGC 6520	&	$-$0.25	&	--	&	--	&	J	&	[57] \\
NGC 3680	&	+0.10	&	0.20	&	--	&	W	&	[10] &	NGC 6611	&	$-$0.46	&	--	&	--	&	J	&	[57] \\
NGC 3680	&	+0.02	&	0.03	&	8	&	D, W	&	[17] &	NGC 6633	&	$-$0.133	&	--	&	39	&	J	&	[7] \\
NGC 3680	&	+0.09	&	0.08	&	25	&	S	&	[41] &	NGC 6633	&	$-$0.01	&	0.03	&	5	&	D	&	[25] \\
NGC 3680	&	$-$0.16	&	0.07	&	8	&	D	&	[49] &	NGC 6633	&	$-$0.02	&	0.05	&	3	&	D	&	[49] \\
NGC 3680	&	$-$0.121	&	0.051	&	8	&	D	&	[61] &	NGC 6649	&	+0.05	&	--	&	--	&	J	&	[57] \\
NGC 3766	&	$-$0.47	&	--	&	--	&	J	&	[57] &	NGC 6705	&	+0.07	&	--	&	52	&	J	&	[7] \\
NGC 3960	&	$-$0.68	&	0.28	&	11	&	W	&	[29] &	NGC 6705	&	$-$0.39	&	--	&	--	&	J	&	[57] \\
NGC 3960	&	$-$0.06	&	0.05	&	3	&	D	&	[49] &	NGC 6716	&	$-$0.311	&	--	&	24	&	J	&	[7] \\
NGC 3960	&	$-$0.170	&	0.131	&	6	&	D	&	[61] &	NGC 6716	&	$-$0.26	&	--	&	--	&	J	&	[57] \\
NGC 4103	&	$-$0.47	&	--	&	--	&	J	&	[57] &	NGC 6755	&	$-$0.03	&	--	&	--	&	J	&	[57] \\
NGC 4349	&	$-$0.23	&	0.08	&	--	&	D, J, W	&	[14] &	NGC 6791	&	$-$0.08	&	0.07	&	7	&	D	&	[30] \\
NGC 4349	&	$-$0.12	&	0.04	&	5	&	D	&	[49] &	NGC 6791	&	+0.50	&	--	&	--	&	J	&	[35] \\
NGC 4349	&	$-$0.060	&	0.123	&	6	&	D	&	[61] &	NGC 6791	&	+0.05	&	0.05	&	--	&	J	&	[43] \\
NGC 4609	&	+0.05	&	0.13	&	1	&	D, W	&	[17] &	NGC 6791	&	+0.00	&	0.14	&	6	&	D	&	[49] \\
\hline
\end{tabular}
\end{center}
\end{table*}

\begin{table*}
\begin{center}
\addtocounter{table}{-1}
\caption[]{continued}
\begin{tabular}{cccccc|cccccc}
\hline
\hline
Cluster	&	[Fe/H]	&	$\sigma$[Fe/H] &	Stars	&	System	&	Ref	&	Cluster	&	[Fe/H]	&	$\sigma$[Fe/H]	&	Stars	&	
System	&	Ref	\\
\hline
NGC 6791	&	+0.094	&	0.141	&	7	&	D	&	[61] &	Pismis 4	&	$-$0.20	&	0.08	&	2	&	D	&	[49] \\
NGC 6791	&	+0.45	&	0.04	&	--	&	C, S	&	[63] &	Pismis 4	&	$-$0.055	&	0.095	&	2	&	D	&	[61] \\
NGC 6823	&	+0.08	&	--	&	--	&	J	&	[57] &	Pismis 20	&	+0.03	&	--	&	--	&	J	&	[57] \\
NGC 6871	&	$-$0.33	&	--	&	--	&	J	&	[57] &	Ruprecht 18	&	+0.02	&	--	&	--	&	D	&	[9] \\
NGC 6913	&	$-$0.55	&	--	&	--	&	J	&	[57] &	Ruprecht 18	&	+0.08	&	--	&	1	&	D	&	[49] \\
NGC 6940	&	+0.014	&	--	&	28	&	J	&	[7] &	Ruprecht 18	&	$-$0.010	&	0.090	&	1	&	D	&	[61] \\
NGC 7044	&	+0.01	&	--	&	--	&	J	&	[57] &	Ruprecht 20	&	$-$0.29	&	0.04	&	4	&	D, W	&	[17] \\
NGC 7082	&	+0.003	&	--	&	52	&	J	&	[7] &	Ruprecht 32	&	$-$0.38	&	--	&	--	&	J	&	[57] \\
NGC 7092	&	+0.01	&	0.01	&	60	&	S	&	[44] &	Ruprecht 46	&	$-$0.22	&	--	&	--	&	D	&	[9] \\
NGC 7209	&	$-$0.01	&	0.11	&	2	&	D	&	[20] &	Ruprecht 46	&	$-$0.12	&	0.09	&	2	&	D	&	[17] \\
NGC 7209	&	$-$0.12	&	--	&	1	&	D	&	[49] &	Ruprecht 46	&	$-$0.13	&	0.18	&	2	&	D	&	[49] \\
NGC 7209	&	+0.070	&	0.050	&	2	&	D	&	[61] &	Ruprecht 79	&	$-$0.14	&	--	&	--	&	J	&	[57] \\
NGC 7209	&	$-$0.07	&	0.03	&	30	&	V	&	[64] &	Ruprecht 97	&	$-$0.59	&	--	&	1	&	D	&	[20] \\
NGC 7235	&	$-$0.62	&	--	&	--	&	J	&	[57] &	Ruprecht 97	&	$-$0.03	&	0.03	&	2	&	W	&	[25] \\
NGC 7380	&	$-$0.42	&	--	&	--	&	J	&	[57] &	Stock 2	&	$-$0.14	&	0.20	&	2	&	D	&	[20] \\
NGC 7419	&	+0.05	&	--	&	--	&	J	&	[57] &	Tombaugh 2	&	$-$0.60	&	0.18	&	--	&	J	&	[43] \\
NGC 7789	&	$-$0.24	&	0.13	&	17	&	D	&	[49] &	Trumpler 1	&	$-$0.71	&	--	&	--	&	J	&	[57] \\
NGC 7789	&	$-$0.089	&	0.152	&	20	&	D	&	[61] &	Trumpler 10	&	$-$0.13	&	0.20	&	--	&	S	&	[37] \\
NGC 7790	&	$-$0.22	&	--	&	--	&	J	&	[57] &	Trumpler 11	&	$-$0.61	&	--	&	--	&	J	&	[57] \\
Pismis 4	&	$-$0.08	&	0.09	&	2	&	D, W	&	[17] &	Trumpler 14	&	$-$0.03	&	--	&	--	&	J	&	[57] \\
\hline
\end{tabular}
\end{center}
\end{table*}
 
\section{A test case from the literature} \label{sect_tadross}

The pitfalls of using two different photometric calibrations for
identical data sets are clearly illustrated in the papers by Tadross (2001, 2003).

This author performed robust derivations of the cluster
age, reddening, distance, and metallicity on the basis of Johnson $UBV$ photometry.
In the first paper, he used average metallicity values derived from the calibrations of
Carney (1979) and Cameron (1985a), whereas in the second, only values from
Cameron (1985a), were applied (Tadross, private communication). The second study was
found to provide far more reliable results compared to the literature. 

Both calibrations are based on the normalized ultraviolet
excess $\delta(U-B)_\mathrm{0.6}$, introduced by Sandage (1969), which is 
compared to spectroscopically
determined elemental abundances. While Carney (1979) used a list of
published abundances of subdwarfs, Cameron (1985a) constructed a grid 
consisting of blanketing lines and lines of constant metallicity based 
on theoretical as well as empirical data. The determined excess values 
were then transformed to [Fe/H] using a second order polynomial in both
references.

Figure \ref{tadross} compares the [Fe/H] values from both Tadross
papers. 
It is obvious that there are seventeen clusters with values of less than $-$1.0\,dex 
and seven with less than $-$1.5\,dex in Tadross (2001), whereas there is only one 
cluster with [Fe/H] less than $-$1.0\,dex in Tadross (2003). We investigated the 
plausibility of these cases of low metallicity.
Using data of Cepheids, Pedicelli et al. (2009) showed that 
[Fe/H] values significant below $-$0.5\,dex are not expected in the Galactic 
disk where open clusters are located, even at distances of 17\,kpc. The 
compilation of open cluster data by Chen et al. (2003) includes only three 
aggregates with $-$0.5\,$>$\,[Fe/H]\,$>$\,$-$1.0\,dex, which is compatible with
the results by Magrini et al. (2009).

\begin{table*}[t]
\begin{center}
\caption[]{The open clusters with very low [Fe/H] estimates from Tadross (2001) and 
the comparison with those from Tadross (2003). The cluster
parameters are based on the list by Paunzen \& Netopil (2006).}
\begin{tabular}{lcccrrcccrrrr}
\hline
\hline
Cluster &	&	[Fe/H]$_{01}$ &	[Fe/H]$_{03}$ &	\multicolumn{1}{c}{$d$} &	\multicolumn{1}{c}{$\sigma$} & N & 
$E(B-V)$ & $\sigma$ & N & \multicolumn{1}{c}{age} & \multicolumn{1}{c}{$\sigma$} & N \\
	&	&	[dex] &	[dex] &	\multicolumn{1}{c}{[pc]} &	\multicolumn{1}{c}{[pc]} & & [mag] & [mag] &
	& [Myr] & [Myr] \\
\hline
Berkeley 7	&	C0150+621 &	$-$1.75	&	$-$0.25 & 2509	& 195	& 3 &	0.80	& 0.00	& 3	& 4	& 1	& 3 \\						
Berkeley 30	&	C0655+032 &	$-$2.67	& +0.10 & 3990	& 1140	& 5	& 0.50	& 0.07	& 5	& 565	& 350	& 5 \\
Berkeley 60	&	C0015+606 &	$-$2.26	&	$-$0.77	&	3245	&	1634	&	2	&	0.62	&	0.35	&	2	&	160	&	3	& 2 \\
Berkeley 62	&	C0057+636 &	$-$2.39	&	$-$0.85	&	2063	&	400	& 6	&	0.84	&	0.04	& 6	&	17	&	12 & 6 \\
NGC 581	&	C0129+604 &	$-$1.54	&	$-$0.85	&	2569	&	229	&	9 &	0.41	&	0.03	& 10 &	21	&	9 & 11 \\
NGC 1750	&	C0500+235 &	$-$1.97	&	$-$0.69	& 646	& 20	& 3	& 0.34	& 0.01	& 3	& 217	& 30	& 3 \\
NGC 2244	&	C0629+049 &	$-$1.77	&	$-$0.46	&	1595	&	119	&	9 &	0.46	&	0.03	&	10 &	5	&	2 &	10 \\
\hline
\end{tabular}
\label{seven}
\end{center}
\end{table*}

In Table \ref{seven}, we list the seven extreme cases together with the cluster parameters
based on the list by Paunzen \& Netopil (2006). For two clusters, Berkeley 7 and Berkeley 30,
respectively, the [Fe/H] determinations from Tadross (2003) seem to be well in the expected range.
For the remaining five clusters, the new determinations are still in the range 
$-$0.46\,$>$\,[Fe/H]\,$>$\,$-$0.85\,dex, which is surprisingly low for the given ages
and distances of the aggregates. For Berkeley 60, a large error in the distance and reddening
from the literature is evident. Ann et al. (2002) list a near solar metallicity (+0.07\,dex) for
this cluster based on isochrone fitting to Johnson $UBVI$ measurements. For NGC 581 (Sanner et al. 1999)
and NGC 2244 (Bonatto \& Bica 2009),
we also find a published solar metallicity. The cluster NGC 1750 is very interesting
in the sense that it overlaps with NGC 1758 (Strai\v{z}ys et al. 2003). Therefore the choice of the 
members is essential to derive correct values. In summary, we find no evidence that
the low metallicities are supported by any other source.

The remaining values (Fig. \ref{tadross}) are within a range of +0.3\,$>$\,[Fe/H]\,$>$\,$-$1.0\,dex, 
which is the expected one for open clusters (Chen et al. 2003). For the following analysis, we intend
to use the discussed metallicities in a proper way because they fill an important gap
within the published investigations. Because the later paper by Tadross supersedes the previous
one, we considered only the values from Tadross (2003) for our further analysis. 

\begin{table*}
\begin{center}
\caption[]{Unweighted averaged [Fe/H] values for 188 Galactic open clusters, 88 of them have more than one measurement.}
\begin{tabular}{cccc|cccc|cccc}
\hline
\hline
Cluster	&	[Fe/H]	&	$\sigma$[Fe/H] &	N	&	Cluster	&	[Fe/H]	&	$\sigma$[Fe/H] &	N	& Cluster	&	[Fe/H]	&	$\sigma$[Fe/H] &	N	\\
\hline
Berkeley 1	&	$-$0.14	&	0.10	&	1	&	NGC 1342	&	$-$0.20	&	0.06	&	2	&	NGC 4349	&	$-$0.13	&	0.08	&	3	\\
Berkeley 2	&	$-$0.59	&	0.10	&	1	&	NGC 1528	&	$-$0.13	&	0.10	&	1	&	NGC 4609	&	+0.05	&	0.13	&	1	\\
Berkeley 7	&	$-$0.25	&	0.10	&	1	&	NGC 1545	&	$-$0.13	&	0.07	&	3	&	NGC 4755	&	$-$0.21	&	0.10	&	1	\\
Berkeley 14	&	$-$1.20	&	0.10	&	1	&	NGC 1662	&	$-$0.11	&	0.08	&	4	&	NGC 4815	&	$-$0.19	&	0.10	&	1	\\
Berkeley 21	&	$-$0.96	&	0.10	&	1	&	NGC 1750	&	$-$0.69	&	0.10	&	1	&	NGC 5138	&	+0.15	&	0.04	&	2	\\
Berkeley 22	&	$-$0.42	&	0.10	&	1	&	NGC 1798	&	$-$0.46	&	0.10	&	1	&	NGC 5168	&	+0.05	&	0.10	&	1	\\
Berkeley 28	&	$-$0.61	&	0.10	&	1	&	NGC 1817	&	$-$0.31	&	0.03	&	2	&	NGC 5281	&	$-$0.18	&	0.10	&	1	\\
Berkeley 29	&	$-$0.30	&	0.10	&	1	&	NGC 1931	&	$-$0.06	&	0.10	&	1	&	NGC 5316	&	+0.11	&	0.10	&	3	\\
Berkeley 30	&	+0.10	&	0.10	&	1	&	NGC 2099	&	+0.14	&	0.06	&	2	&	NGC 5606	&	+0.09	&	0.10	&	1	\\
Berkeley 31	&	$-$0.60	&	0.10	&	1	&	NGC 2112	&	$-$1.30	&	0.22	&	1	&	NGC 5617	&	+0.10	&	0.60	&	2	\\
Berkeley 32	&	$-$0.42	&	0.09	&	3	&	NGC 2158	&	$-$0.29	&	0.09	&	2	&	NGC 5662	&	+0.03	&	0.09	&	2	\\
Berkeley 39	&	$-$0.33	&	0.07	&	3	&	NGC 2168	&	$-$0.27	&	0.12	&	3	&	NGC 5822	&	$-$0.07	&	0.10	&	6	\\
Berkeley 42	&	+0.08	&	0.10	&	1	&	NGC 2192	&	$-$0.22	&	0.10	&	1	&	NGC 6025	&	+0.19	&	0.10	&	1	\\
Berkeley 58	&	+0.09	&	0.10	&	1	&	NGC 2194	&	$-$0.27	&	0.06	&	1	&	NGC 6067	&	+0.08	&	0.07	&	3	\\
Berkeley 60	&	$-$0.77	&	0.10	&	1	&	NGC 2204	&	$-$0.43	&	0.11	&	4	&	NGC 6087	&	$-$0.04	&	0.10	&	1	\\
Berkeley 62	&	$-$0.85	&	0.10	&	1	&	NGC 2232	&	$-$0.20	&	0.21	&	1	&	NGC 6134	&	+0.20	&	0.12	&	5	\\
Berkeley 79	&	$-$0.32	&	0.10	&	1	&	NGC 2236	&	$-$0.30	&	0.20	&	1	&	NGC 6192	&	$-$0.21	&	0.19	&	2	\\
Berkeley 86	&	$-$0.08	&	0.10	&	1	&	NGC 2243	&	$-$0.57	&	0.08	&	3	&	NGC 6204	&	$-$0.14	&	0.10	&	1	\\
Collinder 74	&	+0.05	&	0.10	&	1	&	NGC 2244	&	$-$0.46	&	0.10	&	1	&	NGC 6208	&	$-$0.03	&	0.06	&	1	\\
Collinder 140	&	$-$0.05	&	0.06	&	2	&	NGC 2251	&	$-$0.21	&	0.06	&	2	&	NGC 6231	&	+0.26	&	0.10	&	1	\\
Collinder 173	&	$-$0.06	&	0.20	&	1	&	NGC 2264	&	$-$0.08	&	0.08	&	3	&	NGC 6259	&	+0.04	&	0.04	&	2	\\
Collinder 258	&	$-$0.07	&	0.16	&	3	&	NGC 2266	&	$-$0.26	&	0.20	&	1	&	NGC 6281	&	+0.00	&	0.06	&	4	\\
Collinder 272	&	$-$0.40	&	0.10	&	1	&	NGC 2281	&	$-$0.07	&	0.10	&	1	&	NGC 6405	&	+0.07	&	0.10	&	1	\\
Haffner 8	&	+0.02	&	0.08	&	3	&	NGC 2287	&	$-$0.06	&	0.13	&	5	&	NGC 6425	&	+0.18	&	0.12	&	2	\\
IC 348	&	+0.03	&	0.10	&	1	&	NGC 2301	&	+0.05	&	0.04	&	5	&	NGC 6451	&	$-$0.27	&	0.10	&	2	\\
IC 1311	&	$-$0.23	&	0.10	&	1	&	NGC 2324	&	$-$0.38	&	0.39	&	3	&	NGC 6475	&	+0.04	&	0.03	&	3	\\
IC 1590	&	$-$0.73	&	0.10	&	1	&	NGC 2335	&	+0.08	&	0.15	&	3	&	NGC 6494	&	+0.04	&	0.13	&	3	\\
IC 1805	&	+0.08	&	0.10	&	1	&	NGC 2343	&	$-$0.25	&	0.17	&	2	&	NGC 6520	&	$-$0.25	&	0.10	&	1	\\
IC 2391	&	$-$0.12	&	0.04	&	2	&	NGC 2354	&	$-$0.30	&	0.02	&	1	&	NGC 6611	&	$-$0.46	&	0.10	&	1	\\
IC 2488	&	+0.09	&	0.08	&	1	&	NGC 2355	&	+0.02	&	0.20	&	2	&	NGC 6633	&	$-$0.05	&	0.06	&	3	\\
IC 2602	&	$-$0.23	&	0.10	&	1	&	NGC 2360	&	$-$0.15	&	0.07	&	5	&	NGC 6649	&	+0.05	&	0.10	&	1	\\
IC 2714	&	$-$0.04	&	0.07	&	3	&	NGC 2420	&	$-$0.38	&	0.07	&	4	&	NGC 6705	&	$-$0.10	&	0.36	&	2	\\
IC 4651	&	+0.07	&	0.13	&	10	&	NGC 2422	&	+0.11	&	0.10	&	1	&	NGC 6716	&	$-$0.28	&	0.03	&	2	\\
IC 4665	&	$-$0.11	&	0.07	&	1	&	NGC 2423	&	+0.08	&	0.07	&	4	&	NGC 6755	&	$-$0.03	&	0.10	&	1	\\
IC 4725	&	$-$0.09	&	0.10	&	1	&	NGC 2437	&	$-$0.03	&	0.17	&	2	&	NGC 6791	&	+0.23	&	0.30	&	6	\\
IC 4756	&	+0.01	&	0.03	&	3	&	NGC 2447	&	$-$0.10	&	0.08	&	1	&	NGC 6823	&	+0.08	&	0.10	&	1	\\
IC 4996	&	$-$0.47	&	0.10	&	1	&	NGC 2477	&	$-$0.03	&	0.07	&	4	&	NGC 6871	&	$-$0.33	&	0.10	&	1	\\
King 2	&	$-$0.32	&	0.10	&	1	&	NGC 2482	&	+0.13	&	0.05	&	6	&	NGC 6913	&	$-$0.55	&	0.10	&	1	\\
King 7	&	+0.03	&	0.10	&	1	&	NGC 2489	&	+0.05	&	0.05	&	4	&	NGC 6940	&	+0.01	&	0.10	&	1	\\
King 10	&	+0.09	&	0.10	&	1	&	NGC 2506	&	$-$0.51	&	0.10	&	5	&	NGC 7044	&	+0.01	&	0.10	&	1	\\
King 11	&	$-$0.34	&	0.04	&	2	&	NGC 2516	&	$-$0.07	&	0.16	&	7	&	NGC 7082	&	+0.00	&	0.10	&	1	\\
Loden 807	&	$-$0.18	&	0.05	&	3	&	NGC 2527	&	$-$0.06	&	0.04	&	3	&	NGC 7092	&	+0.01	&	0.01	&	1	\\
Melotte 20	&	+0.04	&	0.03	&	3	&	NGC 2539	&	+0.12	&	0.10	&	5	&	NGC 7209	&	$-$0.03	&	0.08	&	4	\\
Melotte 22	&	+0.06	&	0.01	&	3	&	NGC 2546	&	+0.11	&	0.16	&	4	&	NGC 7235	&	$-$0.62	&	0.10	&	1	\\
Melotte 25	&	+0.14	&	0.04	&	5	&	NGC 2547	&	$-$0.16	&	0.04	&	5	&	NGC 7380	&	$-$0.42	&	0.10	&	1	\\
Melotte 66	&	$-$0.43	&	0.08	&	5	&	NGC 2548	&	+0.08	&	0.07	&	3	&	NGC 7419	&	+0.05	&	0.10	&	1	\\
Melotte 71	&	$-$0.33	&	0.31	&	2	&	NGC 2567	&	+0.03	&	0.14	&	5	&	NGC 7789	&	$-$0.15	&	0.11	&	2	\\
Melotte 105	&	+0.00	&	0.10	&	1	&	NGC 2571	&	+0.05	&	0.10	&	1	&	NGC 7790	&	$-$0.22	&	0.10	&	1	\\
Melotte 111	&	$-$0.03	&	0.03	&	3	&	NGC 2627	&	$-$0.08	&	0.06	&	2	&	Pismis 4	&	$-$0.11	&	0.07	&	3	\\
NGC 103	&	$-$0.85	&	0.10	&	1	&	NGC 2632	&	+0.13	&	0.06	&	6	&	Pismis 20	&	+0.03	&	0.10	&	1	\\
NGC 129	&	$-$0.46	&	0.10	&	1	&	NGC 2658	&	$-$0.94	&	0.10	&	1	&	Ruprecht 18	&	+0.03	&	0.05	&	3	\\
NGC 188	&	$-$0.04	&	0.11	&	3	&	NGC 2660	&	$-$0.27	&	0.32	&	4	&	Ruprecht 20	&	$-$0.29	&	0.04	&	1	\\
NGC 366	&	$-$0.55	&	0.10	&	1	&	NGC 2669	&	$-$0.20	&	0.10	&	1	&	Ruprecht 32	&	$-$0.38	&	0.10	&	1	\\
NGC 381	&	$-$0.04	&	0.10	&	1	&	NGC 2682	&	$-$0.04	&	0.03	&	7	&	Ruprecht 46	&	$-$0.15	&	0.06	&	3	\\
NGC 433	&	$-$0.68	&	0.10	&	1	&	NGC 2818	&	$-$0.30	&	0.02	&	2	&	Ruprecht 79	&	$-$0.14	&	0.10	&	1	\\
NGC 436	&	$-$0.77	&	0.10	&	1	&	NGC 2910	&	$-$0.04	&	0.10	&	1	&	Ruprecht 97	&	$-$0.22	&	0.48	&	2	\\
NGC 457	&	$-$0.46	&	0.10	&	1	&	NGC 2972	&	$-$0.10	&	0.04	&	3	&	Stock 2	&	$-$0.14	&	0.20	&	1	\\
NGC 581	&	$-$0.85	&	0.10	&	1	&	NGC 3114	&	$-$0.05	&	0.08	&	3	&	Tombaugh 2	&	$-$0.60	&	0.18	&	1	\\
NGC 654	&	$-$0.68	&	0.10	&	1	&	NGC 3532	&	$-$0.01	&	0.09	&	3	&	Trumpler 1	&	$-$0.71	&	0.10	&	1	\\
NGC 663	&	$-$0.70	&	0.10	&	1	&	NGC 3680	&	$-$0.02	&	0.11	&	9	&	Trumpler 10	&	$-$0.13	&	0.20	&	1	\\
NGC 752	&	$-$0.15	&	0.11	&	8	&	NGC 3766	&	$-$0.47	&	0.10	&	1	&	Trumpler 11	&	$-$0.61	&	0.10	&	1	\\
NGC 1039	&	$-$0.29	&	0.10	&	1	&	NGC 3960	&	$-$0.23	&	0.29	&	3	&	Trumpler 14	&	$-$0.03	&	0.10	&	1	\\
NGC 1193	&	$-$0.49	&	0.07	&	2	&	NGC 4103	&	$-$0.47	&	0.10	&	1	&		&		&		&		\\		
\hline
\end{tabular}
\label{means}
\end{center}
\end{table*}

\begin{figure}
\begin{center}
\includegraphics[width=85mm]{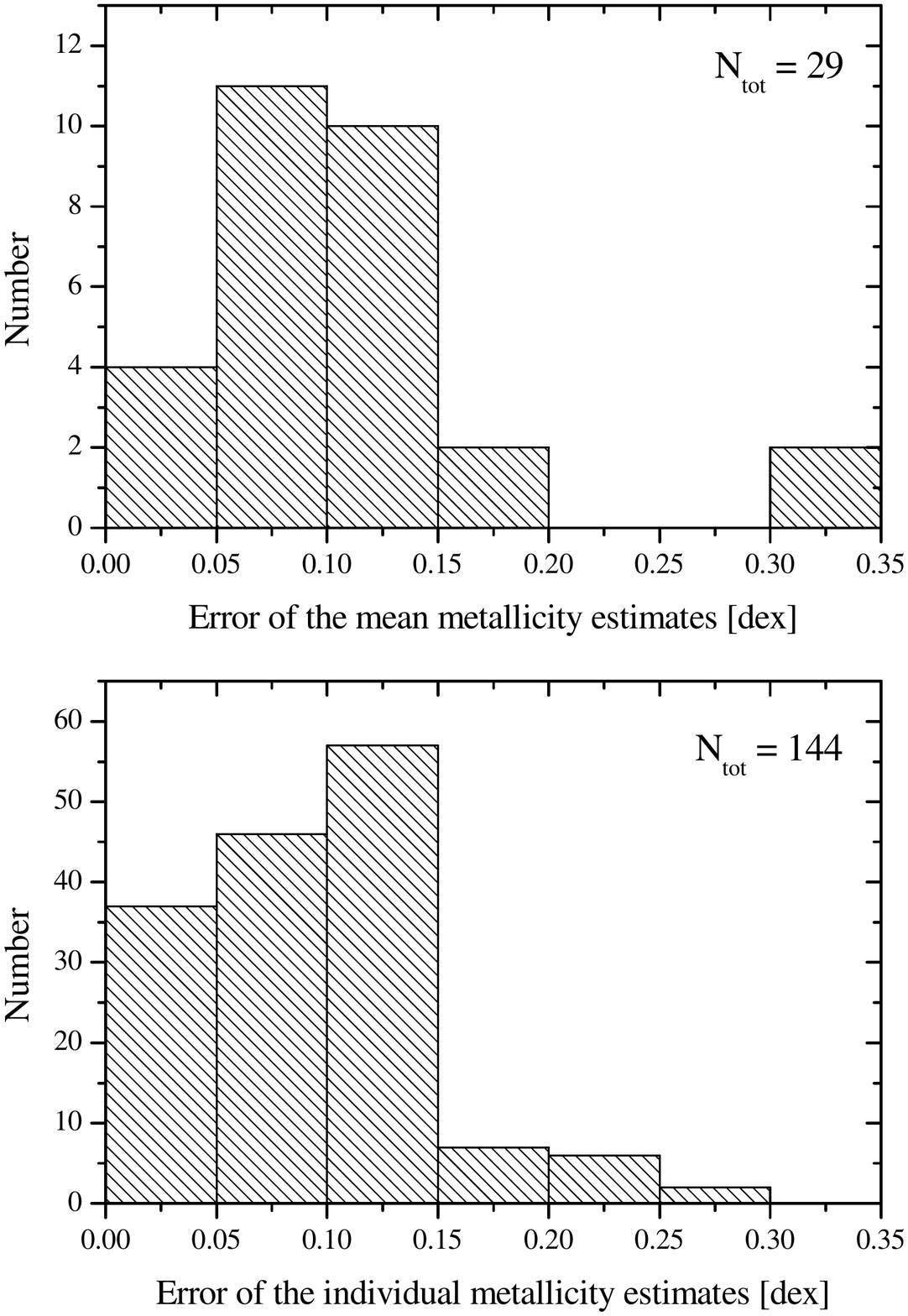}
\caption[]{The error distribution of the individual measurements and means 
for open clusters with N\,$\ge$\,4 estimates from the literature.}
\label{histo}
\end{center}
\end{figure}

\begin{figure}
\begin{center}
\includegraphics[width=85mm]{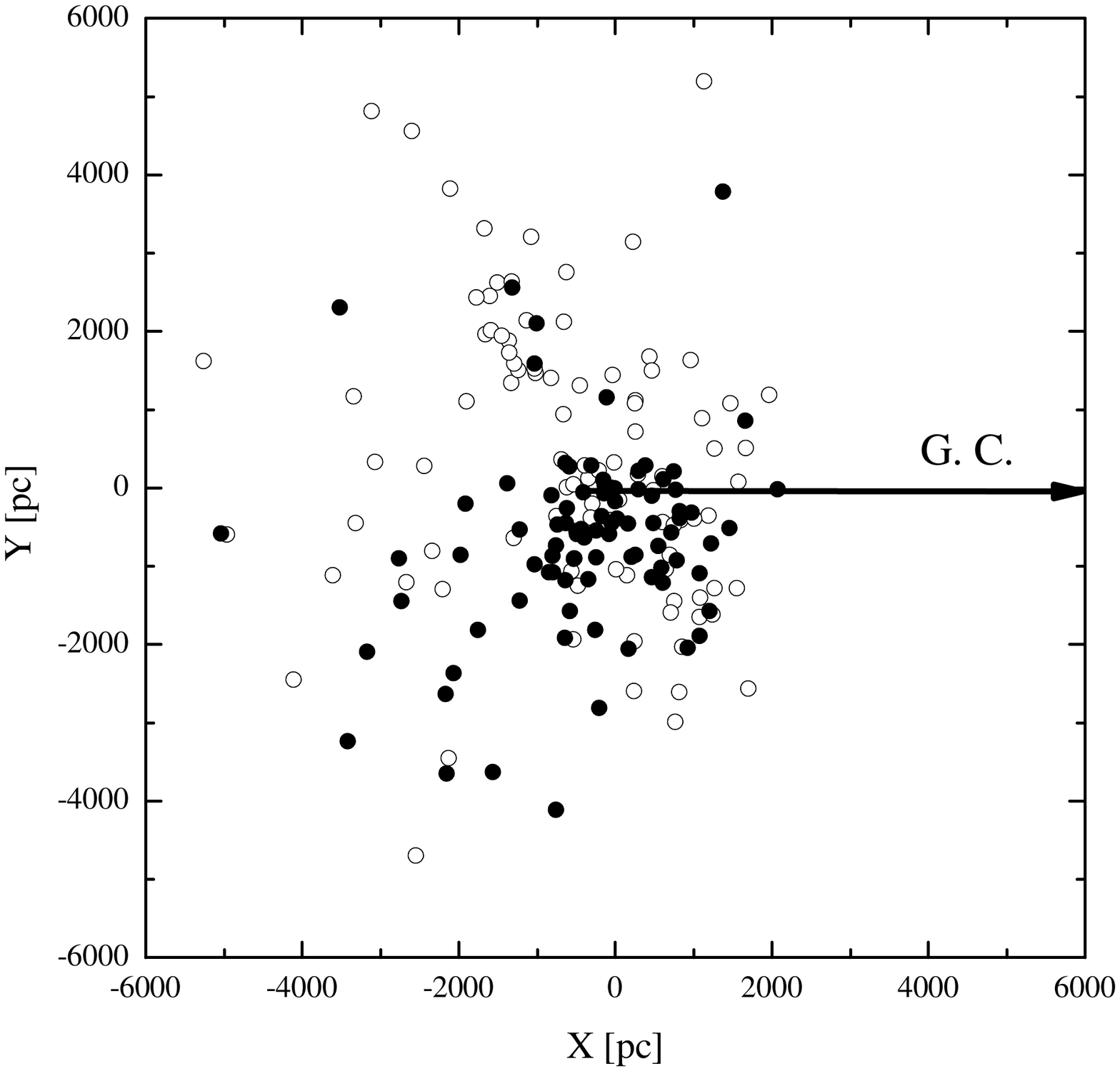}
\caption[]{The Galactic distribution of the final open cluster sample. We excluded 
Berkeley~22, Berkeley~29, and Tombaugh~2 because they are located far away from the 
Sun with a very large distance error. The open circles mark open clusters with only one
available metallicity measurement whereas filled ones are those with more than one value.
The Galactic centre is located at [+8000:0].}
\label{xy_plane}
\end{center}
\end{figure}

\begin{figure}
\begin{center}
\includegraphics[width=85mm,bb= 0 0 1100 900]{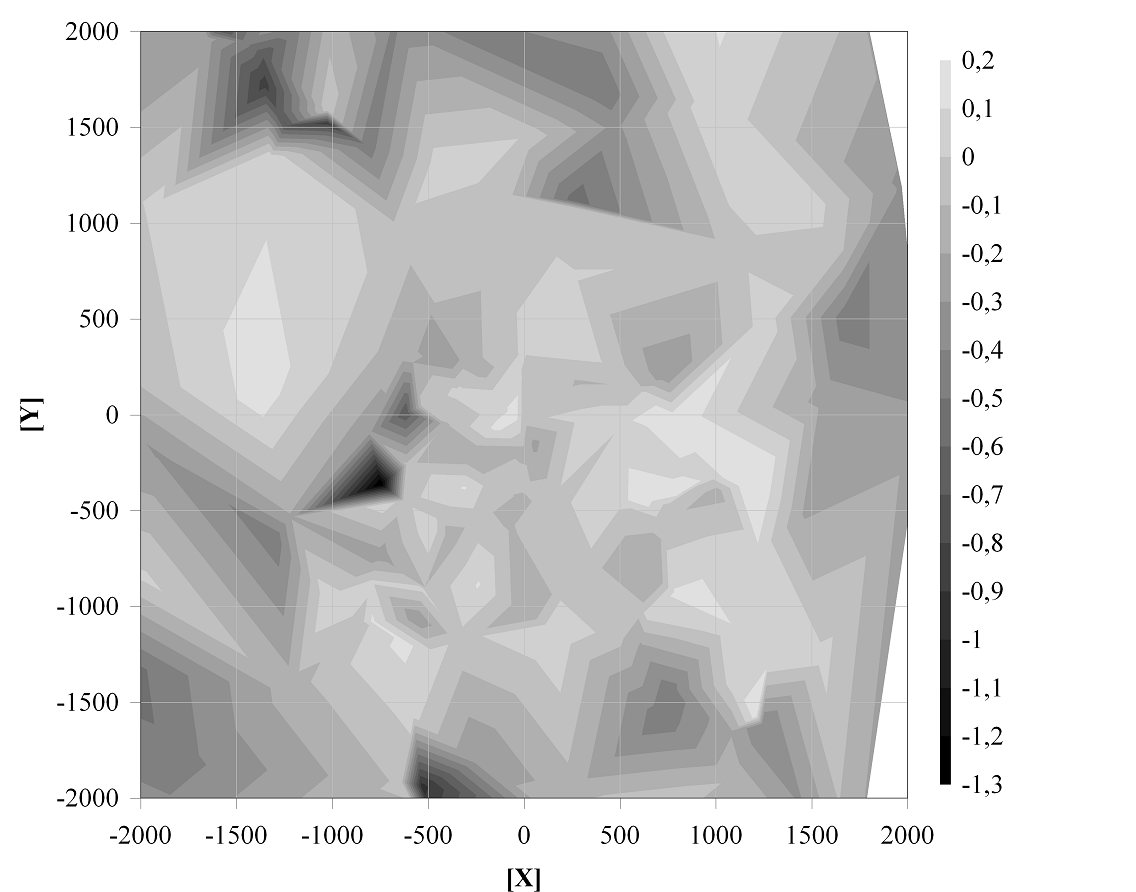}
\caption[]{A contour plot of the metallicity distribution within 
4000$\times$4000\,pc$^{2}$
around the Sun in the [X:Y] plane. The patchy structure is consistent with
results for the Local Bubble which has a diameter of up to 500 pc.}
\label{xy_feh}
\end{center}
\end{figure}

\section{Mean [Fe/H] metallicities and analysis} \label{sect_means}

The basic metallicity calibrations for the individual photometric
systems are, in general, very similar to each other. The starting point is
a set of so-called ``standard objects'' for which the metallicity is clearly 
determined. The effective temperature and the surface gravity are normally
not used in the calibration process.

As a next step, a metallicity sensitive index within a specific photometric
system has to be traced and correlated with the standard set of objects. 
The indices used are listed, in more details, below. There are only very
few examples in which the whole variety of available photometric indices are
used to correlate the metallicity. Martell \& Laughlin (2002) presented 
third order polynomials of Str{\"o}mgren $b-y$, $m_1$, and $c_1$ with the complete 
permutation of these indices resulting in 20 coefficients with a listed precision
of six digits. However, these heuristic calibrations are not widely used.

The metallicity values presented in Table \ref{literature} are based on various
photometric calibrations within six photometric systems listed in Sect.
\ref{ts}. These calibrations are
all based on the individual indices described in the following. 

{\it Caby}: the $hk_0$ index measures the Ca\,II HK line strengths and thus 
the abundances (Anthony-Twarog et al. 1991). The values may be correlated with 
[Fe/H] depending on the temperature sensitive index $(b-y)$.

{\it DDO}: the $\delta CN$ measures the strength of the CN\,4216\AA\,anomaly for red 
giants (Janes 1975). The position of an
object in the $C$(45-48) versus $C$(42-45) diagram is an indicator of the luminosity
for solar metallicity. This location is used to predict the expected value of 
$C$(41-42), a luminosity-dependent metallicity indicator. The $\delta CN$ is then the 
difference between the observed and predicted value that can be transformed via 
a linear relation to [Fe/H]. For example, in the series of papers by Clari{\'a} 
and coworkers, the calibration of Clari{\'a} \& Lapasset (1983) is used until 1991. 
From 1994 on, the calibration of Piatti et al. (1993) is used, based on a new definition of 
the cyanogen anomaly, now designated $\Delta$CN. A comparative test published by Clari{\'a} et al. 
(2003, reference [21] in Table \ref{literature}) indicates that the new calibration can 
result in a metallicity that differs by up to 0.1\,dex from the previous one.

{\it Johnson}: the normalized ultraviolet excess, $\delta (U-B)_{0.6}$, is defined and 
calibrated relative to the Hyades main sequence in the two colour
$(U-B)$ versus $(B-V)$ diagram. It is zero at $(B - V)$\,=\,0.6\,mag for the
metallicity of the Hyades, which is higher than solar, and 
sensitive to the effect of line blanketing (Sandage 1969). 

{\it Str{\"o}mgren}: the $m_1$ index is the standard line blanketing indicator for almost the
complete temperature range (Str{\"o}mgren 1966). 

{\it Vilnius}: $(P-X)$ can be directly correlated with [Fe/H] for cool-type stars
later than G0 (Strai\v{z}ys \& Bartkevi\v{c}ius 1982). 

{\it Washington}: starting with the solar-abundance relations $M-T_1$ versus $C-M$
and using $T_1-T_2$ as a temperature sensitive index, the blanketing in the $M-T_1$
index is correlated with [Fe/H]. The blanketing in the $C-M$ index is correlated
with the CN\,4216\AA\,anomaly of giants (Canterna 1976). Geisler et al. (1991) published 
a new empirical abundance calibration for the Washington system, based on five 
metallicity-sensitive indices, which were used by all subsequent references given 
in Table \ref{literature}.

The standard procedure for deriving the metallicity of a cluster is
always very similar. After a proper selection of cluster members, which is especially 
important in studies using a rather limited number of stars (e.g. red giants), the 
objects have to be dereddened. In principle, the dereddening procedure has to be applied
individually for each star rather than using a mean value, to take into account a 
possible differential reddening. Finally, using the 
respective metallicity calibrations applied to all selected objects, a mean value 
for the aggregate is calculated.

The values from the literature listed in Table \ref{literature} were averaged
and the standard deviation of the mean was calculated. If only one value was available,
the error from the literature was taken. If there is none available, we set the
error to 0.1\,dex, respectively. No weights were introduced in our procedure.
We note that the metallicities
are on a logarithmic scale, so a plain averaging results in a geometric and
not in an arithmetic mean. 

\begin{figure}
\begin{center}
\includegraphics[width=85mm]{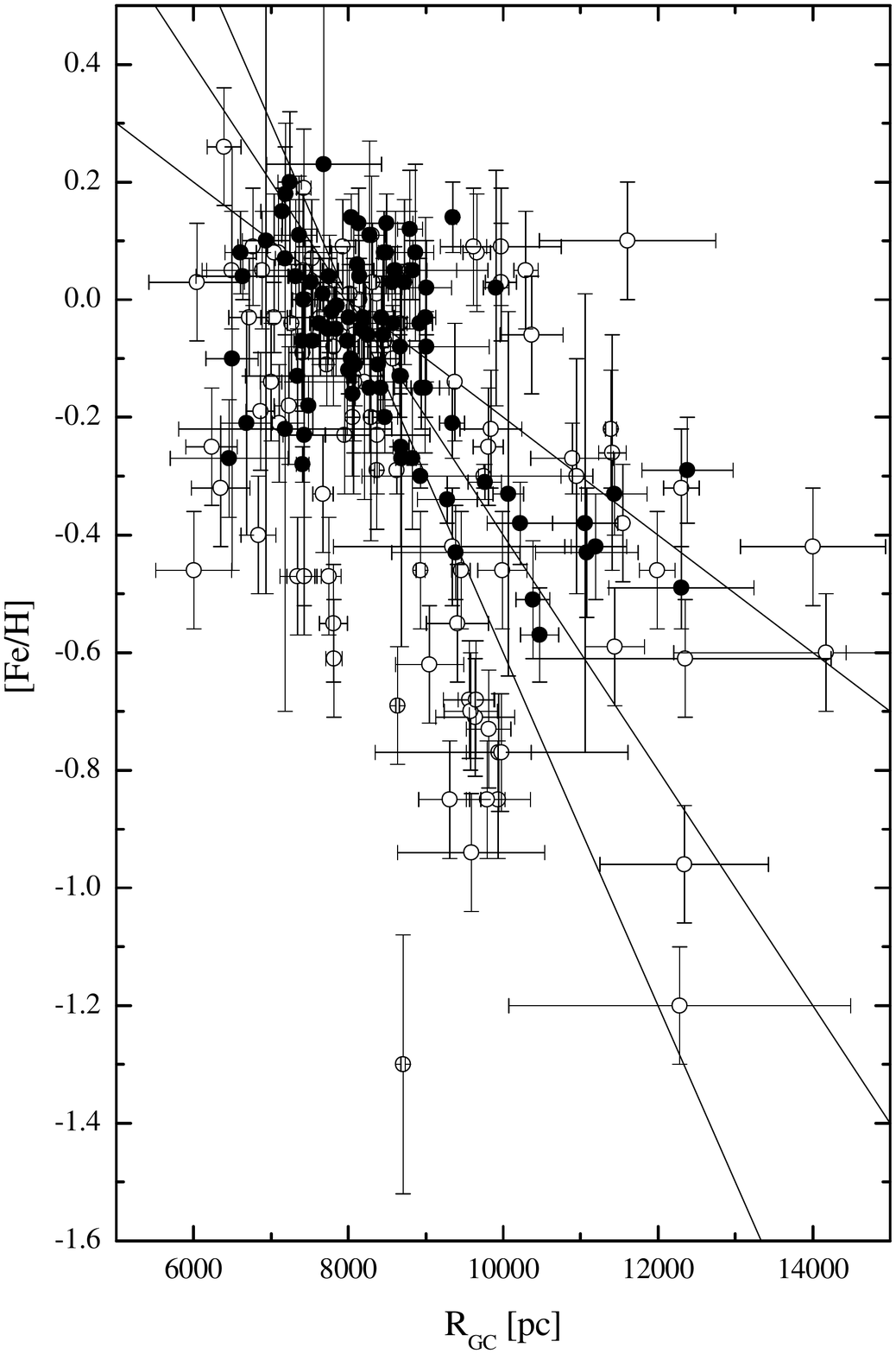}
\caption[]{The [Fe/H] values versus the Galactocentric distance for our sample.
The straight lines represent three different metallicity gradients of 0.1, 0.2, and 
0.3 dex\,kpc$^{-1}$, respectively. The Sun is at $R_{GC}$\,=\,8\,kpc.
The open circles mark open clusters with only one
available metallicity measurement, whereas filled ones are those with more than one value.}
\label{gradient}
\end{center}
\end{figure}

In total, [Fe/H] values for 188 open clusters listed in Table \ref{means} are
available for our further analysis, among which 88 have more than one measurement 
available. The metallicities are comparable to those found for
field Population I type stars in the vicinity of the Sun. 

To test whether the listed errors in the individual metallicity estimates are
realistic, we compared them to the errors in the means for open clusters with 
N\,$\ge$\,4 available values. The latter choice guarantees that no outliers influence
the results. In total, 29 open clusters with 144 individual 
measurements are used. Figure \ref{histo} shows the error distributions for these 
two samples.
We find that 90\% of all values are below 0.15\,dex for both samples. We therefore
conclude that the individual errors are well within the range of those for the means,
and that we do not introduce any bias by calculating the averages from the literature. There are, 
however, a few cases where the errors of the individual measurements might have 
been underestimated, notably NGC 2660 and NGC 6791.

Two of the most interesting and well studied topics are the metallicity distribution
in the vicinity of the Sun and the Galactic metallicity gradient. With the current data set,
we looked at both issues in more detail.

The Galactic distribution within the [X:Y] plane is shown in Fig. \ref{xy_plane}. 
The Galactic centre is located at [+8000:0]. The open clusters Berkeley~22, Berkeley~29, and 
Tombaugh~2 are excluded because they are located far 
away from the Sun with a very large distance error. The distribution of available metallicity
values is not homogeneous within the whole area. The contours were therefore interpolated in a
3D manner using cubic splines. From the plot, it is obvious that the
vicinity of 4000$\times$4000\,pc$^{2}$ can be traced quite reliably. This is an excellent extension to similar 
efforts for the Local Bubble. Lallement et al. (2003) present maps of the density  
distribution of cold gas within $\pm$250 pc from the Sun  
(supplemented by a small number of more distant observations) and  
find a radius for the Local Bubble of 60 pc, with extensions of up to  
250 pc and narrow ``interstellar tunnels'' out to at least 500 pc (Welsh \& Shelton 2009). 
Figure \ref{xy_feh} shows a contour plot of the metallicity distribution 
using our derived mean [Fe/H] values. The patchy structure and the length scale resemble
those of the Local Bubble very well. They continue out to 2000 pc from the Sun.
On the other hand, Luck et al. (2006), investigated a far more extended region of 
20000$\times$10000\,pc$^{2}$, derived from abundances of 54 Cepheids. Their resolution is 
lower than ours but the general trends are comparable.

The metallicity gradient of the Milky Way has been the subject of detailed investigation. 
Observational data of not only star clusters, but also different star groups
such as Cepheids and Population II type stars have been used to place constraints on Galactic
evolutionary models. The published metallicity gradients differ widely. Pedicelli et al. (2009),
for example, concluded that there are two different gradients for the inner and
outer Galactic disc, respectively. Figure \ref{gradient} shows the [Fe/H] abundances 
versus the Galactocentric distance for our sample. The data are consistent with a rather wide range 
of metallicity  gradients (0.1 to 0.3 dex\,kpc$^{-1}$). The estimate depends crucially
on the data points with $R_{GC}$\,$>$\,10\,kpc. But there is an apparent lack of open 
clusters in this region with more than one metallicity estimate. Keeping in mind the patchy structure
of the local environment, the metallicity gradient should be estimated for different lines
of sight within the Galactic plane. This analysis has, to our knowledge, never been done. We will
tackle this topic in the final paper of this series. 

\section{Conclusions and outlook}

We have presented our plans to complete a critical assessment of open  
cluster metallicities. In the first step presented here, we have
collected all metallicity determinations based on photometric  
calibrations.

In total, 406 published values for 188 open clusters 
in 64 references were found that were averaged not taking into account
any weights. A comparison of the papers by Tadross (2001, 2003), which are based on the same data
sets but different applied calibrations, demonstrate how crucial the choice of the
calibration is. The results, sometimes, widely diverge, in the sense that too
low metallicities are derived.

Using published cluster parameters, we investigated the metallicity distribution
in the vicinity of the Sun and the Galactic metallicity gradient in more detail.
We are able to confirm and extend the patchy structure of the Local Bubble out to
2000\,pc from the Sun. The current data set does not allow us to draw any
further conclusions about more distant Galactic regions. For the Galactic metallicity
gradient, no conclusive value can be estimated with the current data set,
because of the patchy local distribution and the lack of very distant open clusters.
Both topics will be investigated in forthcoming studies.

On the basis of the current sample, we propose to search for and analyse elemental
abundances of open cluster members. A detailed comparison of the photometric and
spectroscopic results will essentially help to find the most reliable calibrations.
In addition, apparent off-sets will be able to be identified.

Our last step will be to investigate the influence of the metallicity on 
open cluster parameter estimations that only take into account solar metallicity,
and to provide correction values. 

\begin{acknowledgements}
This work was supported by the financial contributions of the Austrian Agency for International 
Cooperation in Education and Research (WTZ CZ-11/2008, CZ-10/2010 and HR-14/2010), the City of Vienna 
(Hochschuljubil{\"a}umsstiftung project: H-1930/2008), the Forschungsstipendium der 
Universit{\"a}t Wien (F-416), a MOEL grant of the {\"O}FG (Project \#388), a travel grant 
from the Swedish Research Council and the Swedish National Space Board.
This research has made use of the WEBDA 
database, operated at the Institute for Astronomy of the University of Vienna. 
\end{acknowledgements}


\begin{thebibliography}{}
\bibitem[]{} Ann, H. B., Lee, S. H., Sung, H., et al. 2002, AJ, 123, 905
\bibitem[]{} Anthony-Twarog, B. J., \& Twarog, B. A. 2000, AJ, 119, 2282
\bibitem[]{} Anthony-Twarog, B. J., \& Twarog, B. A. 2004, AJ, 127, 1000
\bibitem[]{} Anthony-Twarog, B. J., \& Twarog, B. A. 2006, PASP, 118, 358
\bibitem[]{} Anthony-Twarog, B. J., Twarog, B. A., Laird, J. B., \& Payne, D. 1991, AJ, 101, 1902
\bibitem[]{} Anthony-Twarog, B. J., Atwell, J., \& Twarog, B. A. 2005, AJ, 129, 872
\bibitem[]{} Anthony-Twarog, B. J., Tanner, D., Cracraft, M., \& Twarog, B. A. 2006, AJ, 131, 461
\bibitem[]{} Bessell, M. S. 1995, PASP, 107, 672
\bibitem[]{} Bonatto, C, \& Bica, E. 2009, MNRAS, 394, 2127
\bibitem[]{} Bruntt, H., Frandsen, S., Kjeldsen, H., \& Andersen, M. I. 1999, A\&AS, 140, 135
\bibitem[]{} Cameron, L. M. 1985a, A\&A, 146, 59
\bibitem[]{} Cameron, L. M. 1985b, A\&A, 147, 39
\bibitem[]{} Canterna, R. 1976, AJ, 81, 228
\bibitem[]{} Carney, B. W. 1979, ApJ, 233, 211
\bibitem[]{} Cescutti, G., Matteucci, F., Fran\c{c}ois, P., \& Chiappini, C. 2007, A\&A, 462, 943
\bibitem[]{} Chen, L., Hou, J. L., \& Wang, J. J. 2003, AJ, 125, 1397
\bibitem[]{} Chiappini, C., Matteucci, F., \& Romano, D. 2001, ApJ, 554, 1044
\bibitem[]{} Clari{\'a}, J. J. 1982, A\&AS, 47, 323
\bibitem[]{} Clari{\'a}, J. J. 1985, A\&AS, 59, 195
\bibitem[]{} Clari{\'a}, J. J. 1986, A\&AS, 59, 195
\bibitem[]{} Clari{\'a}, J. J., \& Lapasset, E. 1983, Journal of Astrophysics and Astronomy, 4, 117 
\bibitem[]{} Clari{\'a}, J. J., \& Lapasset, E. 1985, MNRAS, 214, 229
\bibitem[]{} Clari{\'a}, J. J., \& Lapasset, E. 1986a, ApJ, 302, 656
\bibitem[]{} Clari{\'a}, J. J., \& Lapasset, E. 1986b, AJ, 91, 326
\bibitem[]{} Clari{\'a}, J. J., \& Lapasset, E. 1989, MNRAS, 241, 301
\bibitem[]{} Clari{\'a}, J. J., \& Mermilliod, J.-C. 1992, A\&AS, 95, 429
\bibitem[]{} Clari{\'a}, J. J., \& Minniti, D. 1988, The Observatory, 108, 218
\bibitem[]{} Clari{\'a}, J. J., Lapasset, E., \& Minniti, D. 1989, A\&AS, 78, 363
\bibitem[]{} Clari{\'a}, J. J., Lapasset, E., \& Bosio, M. A. 1991, MNRAS, 249, 193
\bibitem[]{} Clari{\'a}, J. J., Mermilliod, J.-C., \& Piatti, A. E. 1999, A\&AS, 134, 301
\bibitem[]{} Clari{\'a}, J. J., Piatti, A. E., \& Osborn, W. 1996, PASP, 108, 672
\bibitem[]{} Clari{\'a}, J. J., Mermilliod, J.-C., Piatti, A. E., \& Minniti, D. 1994, A\&AS, 107, 39
\bibitem[]{} Clari{\'a}, J. J., Piatti, A. E., Lapasset, E., \& Mermilliod, J.-C. 2003, A\&A, 399, 543
\bibitem[]{} Clari{\'a}, J. J., Piatti, A. E., Lapasset, E., \& Parisi, M. C. 2005, Baltic Astronomy, 14, 301
\bibitem[]{} Clari{\'a}, J. J., Piatti, A. E., Parisi, M. C., \& Ahumada A. V. 2007, MNRAS, 379, 159
\bibitem[]{} Clari{\'a}, J. J., Piatti, A. E., Mermilliod, J.-C., \& Palma, T. 2008, AN, 329, 609
\bibitem[]{} Dawson, D. W. 1981, AJ, 86, 237
\bibitem[]{} Dias, W. S., Alessi, B. S., Moitinho, A., \& Lepine, J. R. D. 2002, A\&A, 389, 871
\bibitem[]{} Eggen, O. J. 1983, AJ, 88, 197
\bibitem[]{} Geisler, D. P., \& Smith, V. V. 1984, PASP, 96, 871
\bibitem[]{} Geisler, D. P., Clari{\'a}, J. J., \& Minniti, D. 1991, AJ, 102, 1836
\bibitem[]{} Geisler, D. P., Clari{\'a}, J. J., \& Minniti, D. 1992, AJ ,104 ,1892
\bibitem[]{} Girardi, L., Bressan, L., Bertelli, G., \& Chiosi, C. 2000, A\&AS, 141, 371
\bibitem[]{} Groenewegen, M. A. T., Udalski, A., \& Bono, G. 2008, A\&A, 481, 441
\bibitem[]{} Janes, K. A. 1975, ApJS, 29, 161
\bibitem[]{} Janes, K. A. 1984, PASP, 96, 977
\bibitem[]{} Janes, K. A., \& Smith, G. H. 1984, AJ, 89, 487
\bibitem[]{} Jappsen, A.-K., Glover, S. C. O., Klessen, R. S., \& Mac Low, M.-M. 2007, ApJ, 660, 1332
\bibitem[]{} Kaluzny, J., \& Mazur, B. 1991a, Acta Astronomica, 41, 167
\bibitem[]{} Kaluzny, J., \& Mazur, B. 1991b, Acta Astronomica, 41, 191
\bibitem[]{} Kaluzny, J., \& Mazur, B. 1991c, Acta Astronomica, 41, 279
\bibitem[]{} Kaluzny, J., \& Rucinski, S. M. 1995, A\&AS, 114, 1
\bibitem[]{} Kharchenko, N. V., Piskunov, A. E., R{\"o}ser, S., Schilbach, E., \& Scholz, R.-D. 2005, 
A\&A, 438, 1163
\bibitem[]{} Kharchenko, N. V., Piskunov, A. E., R{\"o}ser, S., et al. 2009, A\&A, 504, 681 
\bibitem[]{} Kyeong, J., Kim, S. C., Hiriart, D., \& Sung, E.-C. 2008, Journal of the Korean 
Astronomical Society, 41, 147
\bibitem[]{} Lallement, R., Welsh, B. Y., Vergely, J. L., Crifo, F., \& Sfeir, D. 2003, A\&A, 411, 447
\bibitem[]{} Luck, R. E., Kovtyukh, V. V., \& Andrievsky, S. M. 2006, AJ, 132, 902
\bibitem[]{} Lynga, G., \& Wramdemark, S. 1984, A\&A, 132, 58
\bibitem[]{} Machida, M. N. 2008, ApJ, 628, L1
\bibitem[]{} Magrini, L., Sestito, P., Randich, S., \& Galli, D. 2009, A\&A, 494, 95
\bibitem[]{} Martell, S., \& Laughlin, G. 2002, ApJ, 577, L45
\bibitem[]{} McClure, R. D., \& van den Bergh, S. 1968, AJ, 73, 313 
\bibitem[]{} McClure, R. D., Twarog, B. A., \& Forrester, W. T. 1981, ApJ, 243, 841
\bibitem[]{} Mermilliod, J.-C., Clari{\'a}, J. J., Andersen, J., Piatti, A. E., \& Mayor, M. 2001, A\&A, 375,
30
\bibitem[]{} Nissen, P. E. 1980, IAUS, 85, 51
\bibitem[]{} Nissen, P. E. 1988, A\&A, 199, 146
\bibitem[]{} Nissen, P. E., Twarog, B. A., \& Crawford, D. L. 1987, AJ, 93, 634 
\bibitem[]{} Noriega-Mendoza, H., \& Ruelas-Mayorgo, A. 1997, AJ, 113, 722
\bibitem[]{} Palous, J., \& Hauck, B. 1986, A\&A, 162, 54
\bibitem[]{} Parisi, M. C., Clari{\'a}, J. J., Piatti, A. E., \& Geisler, D. 2005, MNRAS, 363, 1247
\bibitem[]{} Pastoriza, M. G., \& Ropke, U. O. 1983, AJ, 88, 1769
\bibitem[]{} Paunzen, E., \& Netopil, M. 2006, MNRAS, 371, 1641
\bibitem[]{} Paunzen, E., St{\"u}tz, Ch., \& Maitzen, H. M. 2005, A\&A, 441, 631
\bibitem[]{} Paunzen, E., Maitzen, H. M., Rakos, K. D., \& Schombert, J. 2003, A\&A, 403, 937
\bibitem[]{} Pedicelli, S., Bono, G., \& Lemasle, B. et al. 2009, A\&A, 504, 81  
\bibitem[]{} Philip, A. G. D. 1976, Dudley Obs. Rep., 12, 1
\bibitem[]{} Piatti, A. E., Clari{\'a}, J. J., \& Abadi, M. G. 1995, AJ, 110, 2813
\bibitem[]{} Piatti, A. E., Clari{\'a}, J. J., \& Ahumada A. V. 2003a, MNRAS, 340, 1249
\bibitem[]{} Piatti, A. E., Clari{\'a}, J. J., \& Ahumada A. V. 2003b, MNRAS, 346, 390
\bibitem[]{} Piatti, A. E., Clari{\'a}, J. J., \& Ahumada A. V. 2004, A\&A, 418, 979
\bibitem[]{} Piatti, A. E., Clari{\'a}, J. J., \& Ahumada A. V. 2006, MNRAS, 367, 599
\bibitem[]{} Piatti, A. E., Clari{\'a}, J. J., \& Minniti D. 1993, Journal of Astrophysics and Astronomy, 14, 145
\bibitem[]{} Platais, I., Kozhurina-Platais, V., Barnes, S., et al. 2001, AJ, 122, 1486  
\bibitem[]{} Richtler, T., 1985, A\&AS, 59, 491
\bibitem[]{} Sandage, A. 1969, ApJ, 158, 1115
\bibitem[]{} Sanner, J., Geffert, M., Brunzendorf, J., \& Schmoll, J. 1999, A\&A, 349, 448 
\bibitem[]{} Schaller, G., Schaerer, G., Meynet, G., \& Maeder, A. 1992, A\&AS, 96, 269
\bibitem[]{} Smith, G. H. 1983, PASP, 95, 296
\bibitem[]{} Smith, G. H., \& Hesser, J. E. 1983, PASP, 95, 277
\bibitem[]{} Smriglio, F., Dasgupta, A. K., Boyle, R. P., \& Nandy, K. 1990, A\&A, 228, 399
\bibitem[]{} Strai\v{z}ys, V., \& Bartkevi\v{c}ius, A. 1982,
Vilnius Astronomijos Observatorijos Biuletenis, 61, 22
\bibitem[]{} Strai\v{z}ys, V., Kazlauskas, A., Cerniauskas, A., et al. 2003, Baltic Astron., 12, 323
\bibitem[]{} Str{\"o}mgren, B. 1966, ARA\&A, 4, 433
\bibitem[]{} Sung, H., Bessell, M. S., Lee, B.-W., \& Lee, S.-G. 2002, AJ, 123, 290
\bibitem[]{} Tadross A. L. 2001, New Astronomy, 6, 293
\bibitem[]{} Tadross A. L. 2003, New Astronomy, 8, 737
\bibitem[]{} Twarog, B. A. 1983, ApJ, 267, 207
\bibitem[]{} Twarog, B. A., Anthony-Twarog, B. J., \& Hawarden, T. G. 1995, PASP, 107, 1215
\bibitem[]{} Twarog, B. A., Anthony-Twarog, B. J., \& McClure, R. D. 1993, PASP, 105, 78
\bibitem[]{} Twarog, B. A., Ashman, K. M., \& Anthony-Twarog, B. J. 1997, AJ, 114, 2556
\bibitem[]{} Twarog, B. A., Corder, S., \& Anthony-Twarog, B. J. 2006, AJ, 132, 299
\bibitem[]{} Twarog, B. A., Vargas, L. S., \& Anthony-Twarog, B. J. 2007, AJ, 134, 1777
\bibitem[]{} Vansevicius, V., Platais, I., Paupers, O., \& Abolins, E. 1997, MNRAS, 285, 871
\bibitem[]{} Welsh, B. Y., \& Shelton, R. L. 2009, Ap\&SS, 323, 1
\bibitem[]{} Yong, D., Karakas, A. I., Lambert, D. L., Chieffi, A., \& Limongi, M. 2008, ApJ, 689, 1031
\end{thebibliography}
\end{document}